%% file: acl_latex.tex
\documentclass[11pt]{article}

\usepackage[preprint]{acl}

\usepackage{times}
\usepackage{latexsym}

\usepackage[T1]{fontenc}

\usepackage[utf8]{inputenc}

\usepackage{microtype}

\usepackage{inconsolata}

\usepackage{graphicx}

\usepackage{amsmath,amssymb,amsfonts}
\usepackage{tikz}
\DeclareRobustCommand*\circled[1]{%
  \tikz[baseline=(char.base)]{
    \node[shape=circle,draw,inner sep=0.5pt] (char) {#1};
  }%
}
\usepackage{booktabs}
\usepackage{multirow}
\usepackage{arydshln}
\newcommand{\pair}[2]{%
\begin{tabular}[c]{@{}c@{\;\textbar\;}c@{}}
#1 & #2
\end{tabular}}

\newcommand{\perfhead}{%
\multicolumn{2}{c}{} 
& WER$\downarrow$ basic \textbar{} full
& SS$\uparrow$ basic \textbar{} full
& DNSMOS basic \textbar{} full
& MAE$\downarrow$ basic \textbar{} full \\
\cmidrule(lr){3-3}
\cmidrule(lr){4-4}
\cmidrule(lr){5-5}
\cmidrule(lr){6-6}
}

\newcommand{\subjhead}{%
\multicolumn{2}{c}{} 
& IMOS$\uparrow$
& SMOS$\uparrow$
& PMOS$\uparrow$ \\
\cmidrule(lr){3-3}
\cmidrule(lr){4-4}
\cmidrule(lr){5-5}
}

\usepackage{amssymb}
\usepackage{enumitem}

\title{CosyEdit2: Speech-Editing-Oriented Reinforcement Learning Unlocks Better Zero-Shot TTS}

\author{
\textbf{Junyang Chen\textsuperscript{1}\thanks{Equal contribution}},
\textbf{Yuhang Jia\textsuperscript{1}\footnotemark[1]},
\textbf{Hui Wang\textsuperscript{1}},
\textbf{Jiaming Zhou\textsuperscript{1}},
\\
\textbf{Yongchang Gan\textsuperscript{2}},
\textbf{Yong Qin\textsuperscript{1}\thanks{Corresponding author}}
\\
\\
 \textsuperscript{1}College of Computer Science, Nankai University
 \\
 \textsuperscript{2}College of Artificial Intelligence, Nankai University
\\
 \small{
   \textbf{Correspondence:} \href{chenjunyang@mail.nankai.edu.cn}{chenjunyang@mail.nankai.edu.cn}, \href{qinyong@nankai.edu.cn}{qinyong@nankai.edu.cn}
 }
}

\begin{document}
\maketitle
\begin{abstract}
Speech editing and zero-shot Text-to-Speech (TTS) share a similar generative foundation conditioned on speech prompts, yet speech editing demands far stricter local acoustic consistency with surrounding unedited content. While prior work has shown that Supervised Fine-Tuning (SFT) enables TTS models to acquire functional editing capability, this approach remains fundamentally bottlenecked by imperfect paired editing data and coarse-grained optimization signals. To address these limitations, we propose CosyEdit2, a speech editing model built on a two-stage post-training framework that progresses from supervised editing initialization to editing-oriented Group Relative Policy Optimization (GRPO) over target-speech-free data. Extensive experiments demonstrate that CosyEdit2 not only substantially advances speech editing performance, but also unlocks better zero-shot TTS capability, revealing a deeper mutual relationship between the two tasks. Audio samples are available at \url{https://cjy1018.github.io/CosyEdit2}.
\end{abstract}

\section{Introduction}
Speech editing aims to modify specific regions of an existing utterance according to textual instructions while preserving the semantic coherence and acoustic consistency with the surrounding unedited context. Unlike zero-shot Text-to-Speech (TTS), which primarily targets textual fidelity and speaker similarity, speech editing imposes substantially stricter preservation requirements: the generated content must seamlessly integrate with the surrounding unedited speech in terms of speaker characteristics, prosody, and acoustic environment, leaving no perceptible trace of modification.

Existing speech editing systems can be broadly categorized into cascade and end-to-end paradigms. Cascaded systems decompose the pipeline into forced alignment, edit-span localization, and target speech generation, achieving stable performance but at the cost of complex preprocessing and sensitivity to alignment errors~\cite{jiang2023fluentspeech, peng2024voicecraft}. End-to-end models eliminate explicit alignment by implicitly learning speech-text correspondence during training~\cite{yan2025ming, chen2026cosyedit}, while delivering competitive performance with substantially lower engineering complexity and greater potential for post-training capability elicitation via sequential token modeling.

Despite recent progress, SFT-based speech editing remains fundamentally limited by imperfect supervision and coarse-grained optimization. On the data side, manually constructed paired target recordings used for supervision inevitably contain boundary ambiguity and acoustic inconsistency, directly propagating artifacts into the learned editing behavior~\cite{chen2026cosyedit}. On the optimization side, SFT optimizes speech editing with token-level reconstruction loss, without distinguishing edited from unedited regions or semantic correctness from acoustic preservation. The result is an inherent preservation–accuracy trade-off that defines the performance ceiling of SFT-based approaches.

To overcome these limitations, we propose CosyEdit2, an end-to-end speech editing model built upon a two-stage post-training framework. In Stage 1, SFT initializes the model with basic editing capability. In Stage 2, editing-oriented Group Relative Policy Optimization (GRPO)~\cite{shao2024deepseekmath} is introduced to optimize the model against editing-specific rewards, without requiring any manually constructed target recordings. By replacing imperfect paired supervision with reward-driven fine-grained optimization, GRPO raises the performance ceiling of speech editing while substantially improving zero-shot TTS generalization. We argue that, while CosyEdit unlocked speech editing from TTS, CosyEdit2 not only advances speech editing to a new level of performance, but also unlocks better zero-shot TTS.

Our main contributions are as follows:

\begin{itemize}[itemsep=-1.5pt, topsep=0pt]
    \item We propose a \textbf{target-speech-free} editing data construction approach for GRPO that converts any TTS corpus into editing training data, eliminating the need for manually constructed imperfect target recordings in SFT and enabling precise injection of speech editing capability into pretrained TTS models.
    \item We present the first \textbf{editing-oriented reward} design for speech editing with GRPO, and establish a complete \textbf{post-training framework} instantiated on CosyVoice2, covering SFT-based capability initialization, GRPO-based capability elicitation, and environment-aware vocoder adaptation.
    \item Extensive experiments demonstrate that CosyEdit2 not only achieves superior \textbf{speech editing} performance across multiple benchmarks, but also substantially improves \textbf{zero-shot TTS} capability, revealing a deeper connection between speech editing and synthesis.
\end{itemize}

\section{Related Work}
\subsection{Text-based Speech Editing}
Recent text-based speech editing systems enable localized insertion, deletion, and substitution of spoken content directly through transcript modifications without re-recording. Existing systems mainly follow cascaded or end-to-end paradigms. Cascaded editors first obtain word- or phoneme-level timestamps via speech-text alignment, then synthesize or infill the edited regions. FluentSpeech~\cite{jiang2023fluentspeech} adopts NAR inpainting, while VoiceCraft~\cite{peng2024voicecraft} and SSR-Speech~\cite{wang2025ssr} perform AR neural-codec token infilling. Despite strong performance, these systems depend on explicit alignment and segmentation, making preprocessing complex and exposing generation to alignment errors. End-to-end SLMs instead internalize alignment in unified models. Ming-UniAudio~\cite{yan2025ming} supports instruction-based editing through large-scale speech-language pretraining. CosyEdit~\cite{chen2026cosyedit} offers a lighter alternative by adapting a zero-shot TTS model with task-specific supervised fine-tuning. However, the preservation of unedited regions in such models remains limited.

Multilingual speech editing remains underexplored, as most prior systems focus on English. Recent models like VoiceCraft-X~\cite{zheng2025voicecraft}, LEMAS~\cite{zhao2026lemas}, and Ming-UniAudio attempt multilingual editing, but their editing quality still lags behind mature English systems.

\subsection{RL for Speech Synthesis and Editing}
\paragraph{Speech Synthesis.} Inspired by Reinforcement Learning (RL) in text LLMs, recent work has explored aligning speech generation with human preferences through reward-driven optimization. Early efforts introduced human feedback into zero-shot TTS~\cite{chen2024enhancing} and leveraged self-supervised reverse inference for robustness~\cite{hu2024robust}. SpeechAlign~\cite{zhang2024speechalign} further formalized this paradigm via DPO-style preference alignment on speech reward models. As LLM-based TTS emerged, RL integration became more prominent: CosyVoice2~\cite{du2024cosyvoice2} and GLM-TTS~\cite{cui2025glm} incorporated reward-based fine-tuning into their codec-language pipelines. To address single-reward instability, Multi-Reward GRPO~\cite{zhong2025multi} aggregated multiple reward signals, while differentiable reward optimization~\cite{gao2025differentiable} replaced non-differentiable pipelines with differentiable approximations. 

\begin{figure*}
    \centering
    \includegraphics[width=1\linewidth]{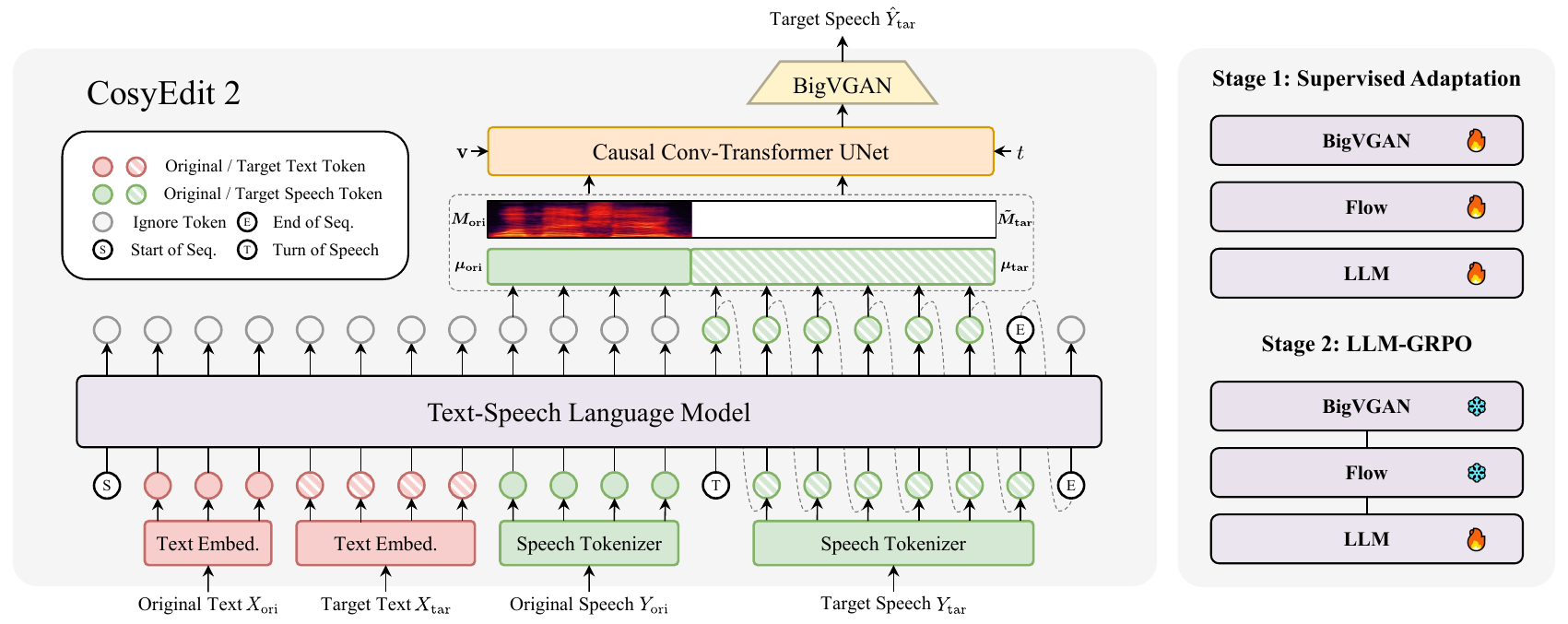}
    \caption{Overview of CosyEdit2. The model reformulates CosyVoice2 for speech editing by conditioning the text-speech language model on original text, target text, and original speech tokens, generating target speech tokens that are decoded by a GOT-CFM Flow and BigVGAN vocoder. The right panel shows the two-stage adaptation: supervised adaptation of LLM, Flow, and BigVGAN respectively, followed by GRPO updating only the LLM.}
    \label{fig:overview}
\end{figure*}

\paragraph{Speech Editing.} Compared with synthesis, the application of RL 
to editing remains relatively underexplored. Recently, ECPA~\cite{ren2026edit} 
applied GRPO to a cascaded speech editing system, demonstrating the promise of 
RL-based optimization for this task. ECPA leverages a pretrained TTS model as 
an implicit critic to optimize semantic-prosodic self-consistency under a TTS 
prior. Our approach differs in that, our editing-oriented GRPO further pursues 
teacher-free, outcome-level optimization specifically tailored to 
speech-editing preferences, directly rewarding both semantic correctness and 
acoustic preservation on decoded speech.

\section{Method}
\subsection{Architecture}
CosyEdit2 adopts the text-speech language modeling backbone of CosyVoice2~\cite{du2024cosyvoice2}, while reformulating its zero-shot prompt-style conditioning interface to speech editing. As illustrated in Figure~\ref{fig:overview}, it consists of text tokenizers, speech tokenizers, an autoregressive text-speech language model, a conditional flow-matching model, and a BigVGAN~\cite{lee2022bigvgan} vocoder. 

To adapt CosyVoice2 for editing, we first reformulate the LLM input interface by separately tokenizing the original and target texts with two identical BPE-based text tokenizers, while representing the original speech and, during training, the target speech with speech tokenizers. We then adopt the GOT-CFM formulation of CosyEdit~\cite{chen2026cosyedit}, where the complete original speech tokens and mel spectrogram are used as global conditions for target speech generation. Finally, we replace the clean-mel-oriented HiFT-GAN~\cite{li2023hiftnet} in CosyVoice2 with a specially trained BigVGAN to better accommodate the diverse acoustic conditions required by speech editing. Detailed module configurations are provided in Appendix~\ref{sec:architecture}.

\subsection{Supervised Adaptation for Speech Editing} 
To elicit speech editing capability from the zero-shot TTS backbone of CosyVoice2, we adopt a two-stage post-training strategy. In the first stage, we perform supervised adaptation on speech editing data, enabling the model to accommodate editing-style inputs that jointly specify the source utterance and the target content, thereby establishing a foundational speech editing capability. We independently adapt the three generation modules:

\paragraph{SFT for LLM and Flow.} For the LLM and Flow modules, we follow the supervised fine-tuning procedures from CosyEdit, training on the 250-hour supervised GigaEdit dataset. Details of these SFT processes and GOT-CFM training in stage 1 are provided in Appendix~\ref{sec:training_details_for_stage_1}.

\paragraph{BigVGAN Training.} For the vocoder, we train BigVGAN to reconstruct waveforms from a mixture of clean and in-the-wild Mel spectrograms. This training exposes the vocoder to both studio-quality speech and recordings with diverse acoustic conditions, enabling more robust waveform generation for speech editing across diverse environments.

\subsection{GRPO for Speech Editing}
In the second stage, we further optimize the language model with GRPO to align generation with editing-specific preferences, including accurate content modification in the edited region and faithful preservation of the unedited regions. Initialized from the Stage-1 trained models, GRPO uses the LLM, Flow, and BigVGAN jointly to produce complete speech-editing rollouts. During this process, the Flow and BigVGAN modules are kept frozen, and only the LLM is updated.

\begin{figure}
    \centering
    \includegraphics[width=1\linewidth]{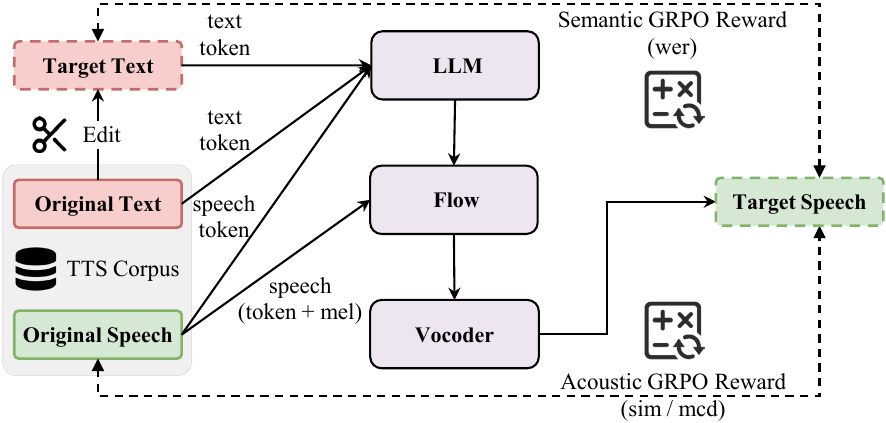}
    \caption{TTS-to-Edit Prompt Construction.}
    \label{fig:tts_to_edit}
\end{figure}

\paragraph{TTS-to-Edit Prompt Construction.} \label{sec:tts_to_edit}
Unlike supervised fine-tuning, this stage does not require manually constructed target speech as supervision. Instead, any TTS-style corpus with speech-transcription pairs can be converted into editing prompts. As illustrated in Figure~\ref{fig:tts_to_edit}, given an utterance and its transcription, we treat them as the original speech and text, and construct the target text by applying various text-level edit operations through rule-based NLP perturbations or LLM-assisted editing. The resulting triplet of original text, target text, and original speech defines an editing prompt.

Conditioned on this prompt, the LLM samples target speech tokens, which are then decoded into waveforms by the fixed Flow and BigVGAN modules. GRPO evaluates the generated speech at the waveform level with editing-specific rewards, rather than imitating a manually constructed target recording at the speech-token level. This avoids supervision artifacts from imperfect edit boundaries or mismatched acoustic conditions, encouraging generations that better satisfy the editing behavior.

\paragraph{Editing-oriented Reward Design.} \label{sec:editing-oriented-reward-design}

For each editing prompt $c=(X_{\mathrm{ori}},X_{\mathrm{tar}},Y_{\mathrm{ori}})$, the policy samples a group of $G$ candidate speech-token sequences $\{Z_i\}_{i=1}^{G}$, which are decoded by the frozen Flow and BigVGAN into waveforms $\{\hat{Y}_{\mathrm{tar}}^{i}\}_{i=1}^{G}$. We evaluate each rollout with three editing-oriented rewards, corresponding to content correctness, acoustic preservation, and speaker consistency. Specifically, the content reward is computed from the word error rate (WER) between the target text and the ASR transcription of the generated speech:
\begin{equation}
\begin{aligned}
w_i &= \mathrm{WER}\!\left(X_{\mathrm{tar}}, \mathrm{ASR}(\hat{Y}_{\mathrm{tar}}^{i})\right), \\
r_i^{\mathrm{wer}} &= \exp\!\left(-k_w \cdot w_i^{\alpha}\right).
\end{aligned}
\end{equation}

The speaker reward directly uses the cosine similarity between speaker embeddings extracted from the original speech and the generated target speech:
\begin{equation}
r_i^{\mathrm{sim}}
=
s_i
=
\frac{
\mathbf{Emb}(Y_{\mathrm{ori}})^{\top}
\mathbf{Emb}(\hat{Y}_{\mathrm{tar}}^{i})
}{
\|\mathbf{Emb}(Y_{\mathrm{ori}})\|_2
\|\mathbf{Emb}(\hat{Y}_{\mathrm{tar}}^{i})\|_2
}.
\end{equation}

Although WER and speaker-similarity rewards are commonly used in GRPO for TTS, optimizing them alone may cause reward hacking, where the model lowers WER at the cost of unnatural word-by-word prosody. We therefore introduce an acoustic preservation reward over the unedited regions. Let $\Omega$ denote the non-edited regions shared by the original and generated speech. We compute
\begin{equation}
\begin{aligned}
m_i &= 
\mathrm{MCD}\!\left(
  \mathrm{DTW}\!\left(Y_{\mathrm{ori}}^{\Omega},\,
  \hat{Y}_{\mathrm{tar}}^{i,\Omega}\right)
\right), \\
r_i^{\mathrm{mcd}}
&=
\exp\!\left(
-k_m \cdot\max(m_i-\delta,0)
\right),
\end{aligned}
\end{equation}
and design a coarse-to-fine, priority-aware reward composition according to speech editing preference. The WER and DTW-aligned mel-cepstral distortion (MCD)~\cite{kubichek1993mel, sakoe1978dynamic} rewards are first combined multiplicatively:
\begin{equation}
r_i^{\mathrm{wer\text{-}mcd}}
=
r_i^{\mathrm{wer}}
\left[
(1-\gamma)+\gamma r_i^{\mathrm{mcd}}
\right],
\end{equation}
where $\gamma$ controls the strength of acoustic-preservation modulation. $r^{\mathrm{wer}}$ serves as a coarse-grained content gate over the whole utterance, measuring whether the sample follows the editing prompt. Given comparable content correctness, $r^{\mathrm{mcd}}$ then selects samples with better fine-grained acoustic preservation in the unedited regions. This ensures the basic editing requirement: correct modification with minimal disturbance. We further add $r^{\mathrm{sim}}$ to rank candidates with better editing quality:
\begin{equation}
r_i
=
\lambda_{\mathrm{c}} r_i^{\mathrm{wer\text{-}mcd}}
+
\lambda_{\mathrm{s}} r_i^{\mathrm{sim}},
\qquad
\lambda_{\mathrm{c}}+\lambda_{\mathrm{s}}=1.
\end{equation}

Here, $\lambda_{\mathrm{c}}$ and $\lambda_{\mathrm{s}}$ balance the editing-reliability reward and the speaker-consistency reward. During training, we dynamically adjust them to emphasize content correctness and acoustic preservation in the early stage, and increase the speaker-similarity weight after the model has learned reliable edits.

\paragraph{GRPO Objective.}
For each prompt, we compute the group-relative advantage by normalizing the rewards within the sampled group:
\begin{equation}
\hat{A}_i =
\frac{r_i-\mu_r}{\sigma_r+\epsilon},
\quad
\mu_r=\frac{1}{G}\sum_{j=1}^{G}r_j .
\end{equation}

\begin{figure}
    \centering
    \includegraphics[width=1\linewidth]{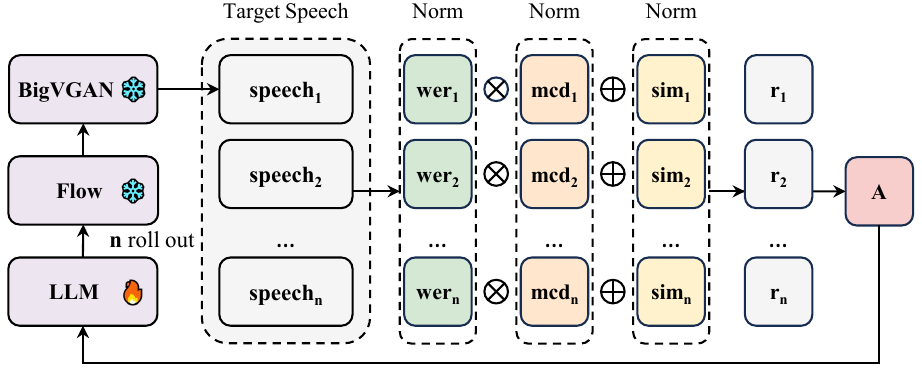}
    \caption{An overview of the editing-oriented GRPO.}
    \label{fig:editing-oriented-GRPO-overview}
\end{figure}

The GRPO objective is then
\begin{multline}
\mathcal{J}_{\mathrm{GRPO}}(\theta)
=
\mathbb{E}
\bigg[
\frac{1}{G}\sum_{i=1}^{G}
\frac{1}{T_i}\sum_{t=1}^{T_i}
\min\!\big[
\rho_{i,t}(\theta)\hat{A}_i,\;
\\
\mathrm{clip}(\rho_{i,t}(\theta),1\pm\epsilon_c)\hat{A}_i
\big]
-
\beta D_{\mathrm{KL}}
\left(\pi_{\theta}\,\|\,\pi_{\mathrm{ref}}\right)
\bigg],
\end{multline}
where
\begin{equation}
\rho_{i,t}(\theta)=\frac{\pi_{\theta}(z_{i,t}\mid c,z_{i,<t})}{\pi_{\theta_{\mathrm{old}}}(z_{i,t}\mid c,z_{i,<t})}
\end{equation}
is the importance ratio. The training loss is $\mathcal{L}_{\mathrm{GRPO}}(\theta)=-\mathcal{J}_{\mathrm{GRPO}}(\theta)$. Here, $\pi_{\theta}$ is the trainable LLM policy, $\pi_{\theta_{\mathrm{old}}}$ is the rollout policy, $\pi_{\mathrm{ref}}$ is the frozen reference policy initialized from the supervised fine-tuned model, and $\epsilon_c$ is the clipping coefficient. Figure~\ref{fig:editing-oriented-GRPO-overview} summarizes this process: 
rewards are computed from decoded waveforms, Flow and BigVGAN 
are used only for rollout, while gradients are applied solely to the LLM.

\section{Experiments}
\subsection{Setup}
\paragraph{Training Data.}
We use separate data for each training stage. The LLM and Flow are first trained on GigaEdit-S from CosyEdit, a 250-hour editing dataset derived from GigaSpeech-S~\cite{chen2021gigaspeech}. BigVGAN is trained on a 625-hour mixture of LibriTTS/LibriTTS-R~\cite{panayotov2015librispeech, koizumi2023libritts} and YODAS2~\cite{li2023yodas}, covering both clean and in-the-wild acoustic conditions. For GRPO, we use only 3,000 randomly sampled utterances from GigaSpeech-XL and synthesize editing prompts via five rule-based perturbation operations using nlpaug~\cite{ma2019nlpaug}, where insertions and substitutions are performed via masked language modeling with RoBERTa~\cite{liu2019roberta}, deletions and swaps via random word removal and reordering, and multi-edits via sequential combinations of all four operations, ensuring precise control over edit type and length. Detailed data construction procedures are in Appendix~\ref{sec:appendix_data_construction_procedures}.

\paragraph{Training Details.}
For GRPO, we initialize the LLM from the Stage-1 checkpoint trained for 8 epochs and keep the Flow and BigVGAN frozen. The LLM is optimized for 380 steps with $G=4$ rollouts per prompt. We set the reward hyperparameters to $k_w=12$, $\alpha=1.5$, $k_m=0.2$, $\delta=2$, and $\gamma=0.5$. The reward weights are scheduled dynamically: $(\lambda_{\mathrm{c}},\lambda_{\mathrm{s}})=(0.9,0.1)$ for the first 290 steps to prioritize content-editing correctness, and $(0.8,0.2)$ for the last 90 steps to strengthen speaker consistency. Rollouts use temperature $0.8$, top-$p=0.95$, top-$k=25$. We train the actor with learning rate $3\times10^{-6}$, KL coefficient $0.001$ and batch size 64, using two NVIDIA H800 GPUs.

\paragraph{Inference.}
CosyVoice2 consists of three independent modules: an autoregressive LLM, a conditional Flow module, and a neural vocoder. This modularity allows us to flexibly compose different task-specific inference pipelines under controlled comparisons. For speech editing, we use the full CosyEdit2 pipeline with the GRPO-optimized LLM, the Stage-1 trained Flow and BigVGAN.  This setting supports both clean recordings and in-the-wild speech with complex backgrounds.

For zero-shot TTS, we replace only the LLM with the GRPO-optimized one and keep the original CosyVoice2 Flow and HiFT-GAN unchanged. This setting follows the zero-shot TTS objective, which prioritizes clean target-speech generation with speaker similarity rather than fully preserving the original acoustic condition, and the original CosyVoice2 acoustic backend is well aligned with this objective. It also isolates the effect of GRPO on the LLM: compared with CosyVoice2, any performance difference mainly reflects the changed language-modeling policy, rather than gains or degradation from a different acoustic backend.

\subsection{Speech Editing}
\paragraph{Evaluation Benchmark.} 
\input{tables/ming-freeform-audio-edit-en}
We evaluate speech editing on the Ming-Freeform-Audio-Edit~\cite{yan2025ming}, which covers insertion, deletion, and substitution operations across English and Chinese basic/full subsets. We report results on the English subset against representative baselines in the main paper. More complete results, including all baseline systems, the Chinese subset, and additional benchmarks, are provided in Appendix~\ref{sec:appendix_speech_editing}.

\paragraph{Baselines.}
We report CosyEdit2 with three representative speech editing systems here: VoiceCraft-X~\cite{zheng2025voicecraft}, SSR-Speech~\cite{wang2025ssr}, and Ming-UniAudio~\cite{yan2025ming}. All baselines are evaluated using their recommended inference configuration. 

\paragraph{Metrics.}
We conduct both objective and subjective evaluations (see 
Appendix~\ref{sec:appendix_subjective_evaluation} for full 
subjective results). Objectively, we report WER for content accuracy, speaker similarity (SS) for speaker preservation, and DNSMOS~\cite{reddy2022dnsmos} for perceptual quality. Prior work often treats higher DNSMOS as better, but it does not imply better editing. In practice, a much higher DNSMOS may indicate implicit denoising or removal of background noise/music rather than faithful preservation. We therefore additionally report DNSMOS mean absolute error ($\mathrm{MAE}_{\mathrm{DNSMOS}}$) between the generated target speech and the original speech, where lower is better, to measure acoustic-quality consistency.

\paragraph{Results.}
Table~\ref{tab:editing_results_en_dnsmos_mae_split} reports the results on the English subset of Ming-Freeform-Audio-Edit. CosyEdit2 consistently outperforms the multilingual cascaded system VoiceCraft-X and the large-scale end-to-end editor Ming-UniAudio across all edit types, and achieves performance comparable to or better than the leading monolingual cascaded system SSR-Speech. Notably, CosyEdit2 obtains the lowest $\mathrm{MAE}_{\mathrm{DNSMOS}}$ across all edit types, indicating better preservation of the original acoustic quality rather than simply generating cleaner speech.

Across edit types, CosyEdit2 performs best on substitution, achieving the lowest WER on both splits and matching the best speaker similarity. For insertion, CosyEdit2 is close to SSR-Speech in WER and SS, while clearly improving $\mathrm{MAE}_{\mathrm{DNSMOS}}$. Deletion remains the most challenging case, where SSR-Speech slightly leads in WER and SS, likely because its explicit speech-text alignment preprocessing simplifies deletion localization. Nevertheless, CosyEdit2 achieves the best acoustic-quality consistency without such external alignment, showing that an end-to-end editing model can approach strong cascaded systems while better preserving the original recording condition.

\subsection{Ablation Experiment}
\input{tables/ablation-experiment}
\paragraph{Evaluation Dataset.}
We conduct ablation studies on RealEdit from VoiceCraft~\cite{peng2024voicecraft}, a 310-sample in-the-wild speech editing benchmark with complex acoustic conditions that better discriminate component-level contributions.

\paragraph{Metrics.}
We report WER, speaker similarity (SS), DNSMOS, and $\mathrm{MAE}_{\mathrm{DNSMOS}}$, as defined in the speech editing evaluation. We additionally report MCD on the unedited 
regions to measure acoustic preservation, computed using pymcd\footnote{\url{https://github.com/chenqi008/pymcd}}.

\paragraph{Results.}
Table~\ref{tab:ablation_realedit} shows the ablation results on RealEdit. We use CosyVoice2 in its zero-shot TTS mode for speech editing as the same-backbone baseline. Although CosyVoice2 achieves the lowest WER, it performs much worse in MCD and $\mathrm{MAE}_{\mathrm{DNSMOS}}$, indicating weak preservation of the original in-the-wild acoustic condition. This is expected because zero-shot TTS does not explicitly preserve non-edited regions and tends to generate cleaner studio-like speech, which can be easier for ASR but less faithful to the original recording. Our case analysis further shows that the higher WER of CosyEdit2 mainly comes from ASR errors caused by preserved background noise or complex prosody, rather than semantic editing errors.

Although SFT improves acoustic preservation over CosyVoice2, yielding better SS and MCD, it severely degrades content accuracy, increasing WER from 4.14 to 5.83, revealing an inherent preservation--accuracy trade-off under imperfect, coarse-grained supervision. Editing-oriented GRPO breaks this trade-off, reducing WER from 5.83 to 4.71 while further improving both SS and MCD. The adapted Flow further improves preservation, reducing MCD from 5.50 to 4.07 and $\mathrm{MAE}_{\mathrm{DNSMOS}}$ from 0.210 to 0.134. Compared with the original HiFT-GAN vocoder, BigVGAN improves waveform reconstruction in the full system, leading to better SS, MCD, and $\mathrm{MAE}_{\mathrm{DNSMOS}}$. Overall, the full CosyEdit2 pipeline achieves the best SS, MCD, and acoustic-quality consistency while maintaining competitive content accuracy.

\subsection{Zero-Shot TTS}
\paragraph{Evaluation Benchmark.}
We evaluate zero-shot TTS on CV3-EVAL, derived from the CosyVoice3 evaluation suite~\cite{du2025cosyvoice3}, which covers multilingual and cross-lingual zero-shot TTS scenarios. For CV3-EVAL, most prompt utterances contain long leading or trailing non-speech regions, such as silence or noise. Since CosyEdit2 relies on model-internal implicit speech-text alignment and is designed to strictly preserve acoustic conditions from the input speech, these regions may be inherited as prompt style cues, which are suitable for speech editing but undesirable for standard zero-shot TTS. We therefore apply VAD-based trimming~\cite{silero2024vad} to the prompt speech before inference and use the same preprocessing for all baselines. Additional SEED-TTS-EVAL~\cite{anastassiou2024seed} results are provided in Appendix~\ref{sec:appendix_seed_tts_eval}.

\input{tables/cv3-eval-simple}
\input{tables/cv3-eval-hard}
\input{tables/cv3-eval-coss-lingual}

\paragraph{Baselines.}
We compare CosyEdit2 with VoiceCraft-X~\cite{zheng2025voicecraft}, SSR-Speech~\cite{wang2025ssr}, CosyEdit~\cite{chen2026cosyedit}, and CosyVoice2~\cite{du2024cosyvoice2} as the same-backbone baseline. All baselines are run with their official checkpoints and inference configurations, and the same data preprocessing is applied to ensure a fair comparison.

\paragraph{Metrics.}
Following the official CV3-EVAL protocol, we report WER/CER for content intelligibility, speaker similarity (SS) for voice cloning fidelity, and DNSMOS for speech quality. Subjective evaluation results are provided in 
Appendix~\ref{sec:appendix_subjective_evaluation}.

\paragraph{Results.}
Tables~\ref{tab:cv3_eval_multilingual_vc}--\ref{tab:cv3_eval_cross_lingual_zeroshot_prompt_lang} report the zero-shot TTS results on CV3-EVAL. CosyEdit2 achieves the lowest error rates across multilingual voice cloning, hard samples, and cross-lingual voice cloning. On the multilingual subset, it improves over the same-backbone CosyVoice2 baseline in every language, with especially clear gains in Japanese and Korean. On the hard subset, CosyEdit2 reduces hard-zh CER from 15.70 to 8.06 and hard-en WER from 8.11 to 5.93 while maintaining comparable or better SS and DNSMOS. These improvements primarily emerge during the GRPO stage rather than from supervised adaptation alone, as performance drops significantly without GRPO. On the cross-lingual subset, CosyEdit2 outperforms all baselines across all target-prompt language pairs. These results suggest that editing-oriented GRPO improves more than speech editing itself: the learned optimization transfers effectively to zero-shot TTS, generalizes across multilingual and cross-lingual settings, and remains robust under challenging scenarios.

\subsection{Discussion}
To understand why a speech editing model can improve zero-shot TTS, we interpret zero-shot TTS as a special case of speech editing under a unified conditional speech generation view. Specifically, when performing zero-shot TTS, the model treats the entire target utterance as the editing region and solves the problem as a full-tail insertion or complete content replacement task. Under this formulation, both tasks share the same core capability requirement: understanding contextual conditions, preserving speaker-related acoustic cues, and generating speech that faithfully follows textual instructions---fundamentally, a form of speech prompt-conditioned in-context learning.

Consequently, editing-oriented GRPO training inherently boosts this in-context learning capability. Semantically, the reward encourages stronger speech-text alignment, reducing hallucination-induced omissions and repetitions in the generated content. Acoustically, the requirement to reconstruct unedited regions compels the model to more meticulously leverage speaker characteristics and other acoustic cues from the prompt speech. Furthermore, it enhances fine-grained articulatory clarity beyond coarse-grained semantic correctness. This gain is particularly evident in the hard subset of CV3-EVAL, which comprises tongue-twister-like sentences, repeated words, and lengthy utterances. Our case studies show that errors stemming from omissions, insertions, and mispronunciations are substantially reduced after GRPO.

Remarkably, although trained solely on English datasets, editing-oriented GRPO evades catastrophic forgetting and instead yields consistent multilingual and cross-lingual gains. We attribute this transfer to a shared mechanism: GRPO strengthens the in-context learning capability underlying prompt-conditioned speech generation, rather than adapting to language-specific patterns, thereby enabling generalization to unseen languages.

\section{Conclusion}
We present CosyEdit2, a speech editing model built on a two-stage post-training framework that bridges speech editing and zero-shot TTS. Stage 1 leverages a pretrained zero-shot TTS model to bootstrap speech editing, exploiting its inherent voice cloning capability for initialization. Stage 2 introduces editing-oriented GRPO to overcome the limitations of SFT caused by imperfect paired supervision and coarse-grained optimization signals. Experiments show that CosyEdit2 not only advances speech editing performance but also feeds back into zero-shot TTS, and even generalizes across languages. These findings suggest that editing-oriented GRPO can strengthen the shared in-context learning capability underlying prompt-conditioned speech generation, revealing a deeper bidirectional connection between editing and synthesis.

\section*{Limitations}
First, the design space of editing-oriented GRPO remains underexplored. Our current reward formulation and hyperparameter settings are derived from task understanding and iterative human listening during training. Although this setup already yields substantial gains, more fine-grained reward formulations that separately model edited and unedited regions, alternative aggregation strategies, and adaptive weighting mechanisms may further improve optimization stability and editing fidelity.

Second, the language coverage of our framework is fundamentally constrained by the underlying pretrained TTS model. CosyEdit2 is built upon CosyVoice2, which supports only Chinese, English, Japanese, and Korean. Although our method exhibits encouraging cross-lingual generalization, these languages still cover only a limited portion of global linguistic diversity. Extending the framework to newer and stronger multilingual TTS backbones and to low-resource languages or dialects remains an important long-term direction.

Finally, our current framework mainly focuses on speech content editing. Benefiting from the pretrained tokenizer and large-scale training data of CosyVoice2, CosyEdit2 partially inherits the ability to generate paralinguistic acoustic events such as laughter, breathing, coughing, and sighs. However, broader acoustic editing capabilities, including emotion conversion, pitch manipulation, speaking-style control, and other fine-grained prosodic modifications, remain insufficiently explored.

\section*{Ethical Considerations}
CosyEdit2 enables high-fidelity speech editing and zero-shot voice generation from short prompt speech, which inevitably raises concerns regarding misuse. Similar to other advanced speech editing systems, the model could potentially be abused for unauthorized voice impersonation, deceptive content creation, misinformation propagation, or other malicious applications involving synthetic audio.

The risks are further amplified by the strong acoustic preservation capability of speech editing models. Unlike conventional zero-shot TTS, speech editing can retain much of the original recording environment, prosody, and speaking characteristics while modifying only partial content, making edited audio potentially more difficult for humans to distinguish from authentic recordings.

Our work is intended solely for legitimate and beneficial applications, such as speech correction, accessibility support, multimedia production, and human--computer interaction research. We do not encourage or endorse any use of the technology for impersonation, fraud, harassment, misinformation, or other harmful purposes. To promote responsible research, we emphasize that future deployment of such systems should be accompanied by appropriate safeguards, including consent-aware usage policies, watermarking or synthetic-audio detection techniques, and careful human oversight in high-stakes scenarios.

At the same time, we believe that open research on speech editing remains important for advancing both capability and safety. Studying these systems in an open academic setting can help the community better understand their risks, develop more reliable detection and attribution methods, and establish responsible norms for future prompt-conditioned speech generation technologies.

\bibliography{custom}

\clearpage

\appendix
\section{A Unified Perspective on Zero-Shot TTS and Speech Editing}
\label{sec:unified_view}

\subsection{Task Formulation}
Zero-shot TTS and speech content editing can both be formulated as conditional speech generation. In zero-shot TTS, given a prompt speech $Y_{\mathrm{p}}$ and a target text $X_{\mathrm{tar}}$, the model generates a target speech $\hat{Y}_{\mathrm{tar}}$ that follows the target linguistic content while preserving the speaker identity of the prompt. Speech editing takes a more constrained form: given the original speech $Y_{\mathrm{ori}}$, the original text $X_{\mathrm{ori}}$, and the target text $X_{\mathrm{tar}}$, the model generates an edited speech $\hat{Y}_{\mathrm{tar}}$ that modifies the intended content while preserving the remaining parts of the original utterance.

From this perspective, zero-shot TTS can be regarded as a full-replacement or full-tail insertion case of speech editing, where the entire linguistic content is regenerated from the target text. Speech editing, in contrast, is a localized replacement problem: only the edited region should change, while the unedited regions should remain consistent with the original speech.

\subsection{Preservation Requirement}
The key distinction between the two tasks lies in the scope of preservation. Standard zero-shot TTS mainly optimizes target text rendering and speaker identity matching. Although modern zero-shot TTS systems may implicitly retain prompt-level prosody or recording conditions~\cite{chen2025neural, du2024cosyvoice, zhou2026indextts2}, such properties are typically not explicitly constrained at the level required for localized editing.

Speech editing requires a stronger preservation objective. The model should modify only the regions specified by the textual edit while maintaining consistency in both linguistic content and acoustic conditions elsewhere. This includes speaker timbre, prosody, background noise, reverberation, and recording characteristics. When preservation is insufficient, the edited speech may resemble independently synthesized TTS outputs rather than a seamless continuation of the original utterance. Existing supervised objectives provide limited direct optimization signals for such holistic preservation, motivating our editing-oriented training objective that jointly optimizes content correctness, speaker consistency, and acoustic preservation.

\subsection{Spectrogram Comparison}
\begin{figure*}[t]
    \centering
    \includegraphics[width=1\linewidth]{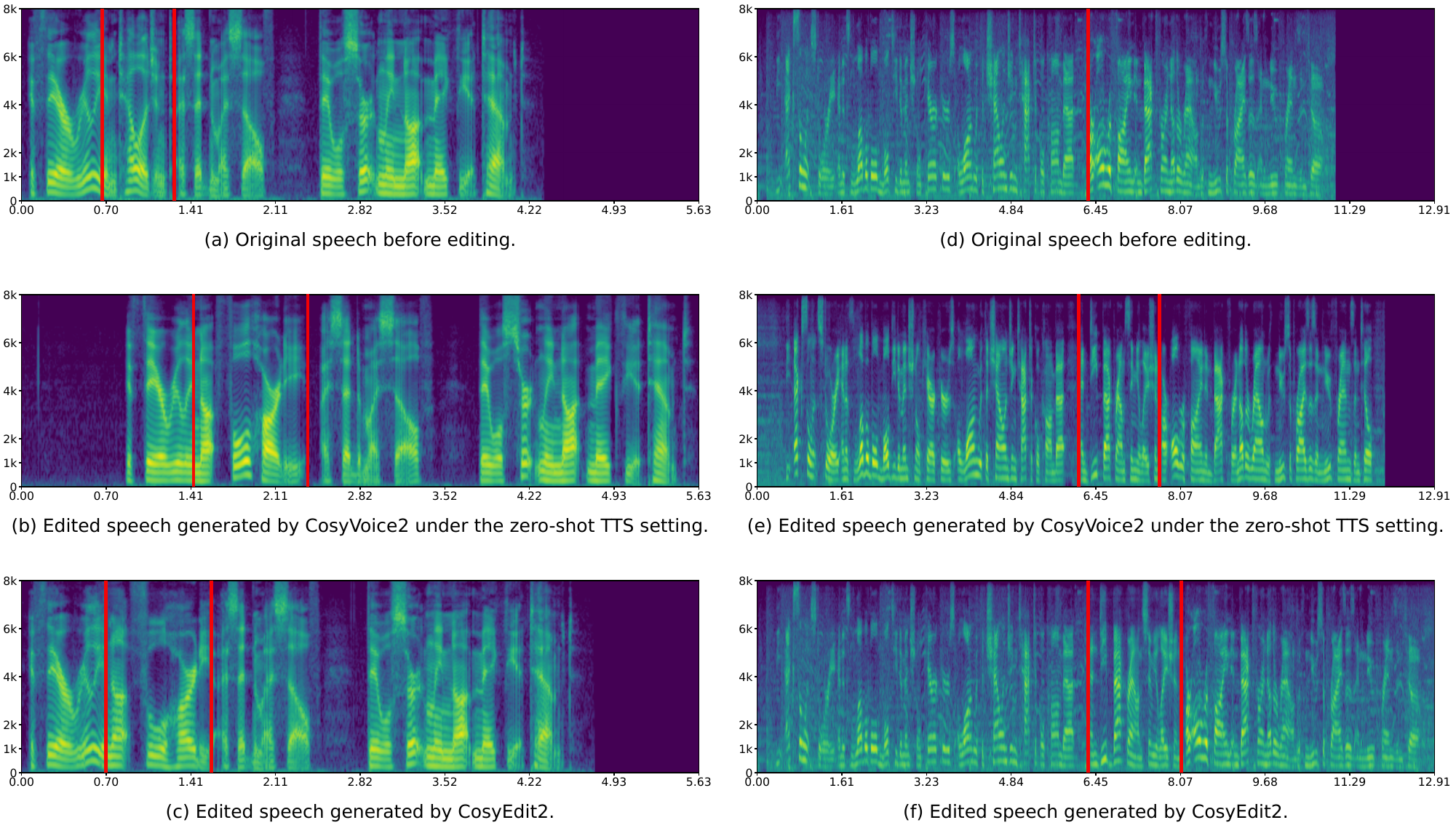}
    \caption{Spectrogram comparison between zero-shot TTS and speech editing. The region between the two red vertical lines indicates the edited segment. The left column~(a--c) illustrates a \textit{substitution} task, while the right column~(d--f) illustrates an \textit{insertion} task.}
    \label{fig:zeroshot_tts_vs_speech_editing}
\end{figure*}

Figure~\ref{fig:zeroshot_tts_vs_speech_editing} illustrates the practical difference between the two tasks. In the substitution case~(a--c), the non-edited regions of the zero-shot TTS output~(b) show clear temporal misalignment. In the insertion case~(d--f), zero-shot TTS output~(e) exhibits degraded preservation particularly in high-frequency components. In contrast, CosyEdit2 outputs~(c) and~(f) better preserve both prosodic contour and spectral detail in the non-edited regions.

This distinction is crucial for evaluating speech editing systems. A model that only produces the correct target text is not necessarily a good editor if it changes the non-edited regions. Therefore, beyond intelligibility and speaker similarity, speech editing requires explicit measurement and optimization of preservation quality, which is the central motivation behind our reward design and acoustic-consistency evaluation.

\section{Architecture Details}
\label{sec:architecture}
\subsection{Text Tokenizer}
CosyEdit2 encodes textual conditions using BPE-based tokenizers, following CosyVoice2~\cite{du2024cosyvoice2} in bypassing an explicit phoneme frontend to allow end-to-end learning of contextual pronunciation patterns. To accommodate the speech editing task, we use two identical text tokenizers to separately encode the original text $X_{\mathrm{ori}}$ and the target text $X_{\mathrm{tar}}$. The two token streams are then sequentially concatenated to form the LLM input sequence, allowing the language model to learn the edit operation implicitly from the contrast between the original and target texts.

\subsection{Speech Tokenizer}
The speech tokenizer follows the supervised semantic token design of CosyVoice2. It extracts discrete speech tokens from waveform inputs using an ASR-oriented encoder with finite scalar quantization (FSQ), producing semantic speech tokens at a low frame rate. In CosyEdit2, the original speech $Y_{\mathrm{ori}}$ is tokenized as part of the editing condition, while the target speech $Y_{\mathrm{tar}}$ is tokenized during supervised training as the prediction target for the LLM. This allows the LLM to model speech editing as conditional semantic-token generation.

\subsection{Unified Text-Speech Language Model}
In CosyVoice2, the language model is an autoregressive unified text-speech model built on the Qwen2.5-0.5B~\cite{hui2024qwen2}. The LLM predicts supervised semantic speech tokens from text and prompt conditions with next-token prediction. CosyEdit2 extends this interface by providing the original text, target text, and original speech tokens as joint conditions. The input sequence contains special tokens indicating the start of sequence, turn of speech, and end of sequence. During inference, the LLM autoregressively generates target speech tokens conditioned on the editing prompt. In the GRPO stage, only this LLM is updated, while downstream acoustic modules remain frozen.

\subsection{Conditional Flow-matching Model}
The Flow module converts semantic speech tokens into Mel spectrograms. To make it suitable for editing, CosyEdit2 adopts the GOT-CFM formulation of CosyEdit~\cite{chen2026cosyedit}. Specifically, the complete original speech tokens and the original Mel spectrogram are used as global conditions for target speech generation. This conditioning design provides the Flow with both semantic and acoustic context from the full original utterance, helping it preserve unedited regions.

\subsection{BigVGAN Vocoder}
The vocoder reconstructs waveform audio from the generated Mel spectrogram. CosyVoice2 originally uses a HiFT-GAN~\cite{li2023hiftnet} vocoder for clean zero-shot TTS synthesis, which operates as a fast frequency domain variant inherited from the classic HiFi-GAN \cite{kong2020hifi} framework. CosyEdit2 replaces it with BigVGAN~\cite{lee2022bigvgan}, a GAN-based universal vocoder with periodic activation functions and anti-aliased representations. These architectural designs provide a stronger inductive bias for high-fidelity waveform generation and improve robustness to diverse recording conditions. This makes CosyEdit2 better suited to speech editing, which often demands modeling more complex and diverse acoustic conditions than clean TTS synthesis.

\section{Training Details for Stage 1}
\label{sec:training_details_for_stage_1}
\subsection{LLM Training}
We first adapt the text-speech language model with supervised learning. Given the original text $X_{\mathrm{ori}}$, target text $X_{\mathrm{tar}}$, original speech $Y_{\mathrm{ori}}$, and target speech $Y_{\mathrm{tar}}$, we encode the texts with BPE-based text tokenizers and extract discrete speech tokens from the original and target speech:
\begin{align}
\mu_{\mathrm{ori}} &= \mathrm{Tok}_{s}(Y_{\mathrm{ori}}), \\
\mu_{\mathrm{tar}} &= \mathrm{Tok}_{s}(Y_{\mathrm{tar}}).
\end{align}

Unlike the inference input sequence organization of zero-shot TTS in CosyVoice2, CosyEdit2 treats the original speech tokens as part of the editing condition and uses a turn-of-speech token $\circled{T}$ to separate the conditioning speech from the autoregressively generated target speech. The LLM input is organized as
\begin{equation}
[\circled{S},\ X_{\mathrm{ori}},\ X_{\mathrm{tar}},\ \mu_{\mathrm{ori}},\ \circled{T}],
\end{equation}
and the model predicts $\mu_{\mathrm{tar}}$ until the end-of-sequence token $\circled{E}$ is generated. This design explicitly separates the source speech from the target speech-token generation, encouraging the LLM to learn editing-oriented speech-text alignment while preserving source-side acoustic cues.

Formally, with the editing condition
$c=(X_{\mathrm{ori}},X_{\mathrm{tar}},\mu_{\mathrm{ori}})$, the LLM is optimized with the next-token prediction loss over the target speech-token sequence:
\begin{equation}
\begin{aligned}
\mathcal{L}_{\mathrm{LM}}
= &-\frac{1}{T_{\mathrm{tar}}+1} \\
  &\sum_{t=1}^{T_{\mathrm{tar}}+1}
   \log p_{\theta}
   \left(\bar{\mu}_{\mathrm{tar},t} \mid c,\bar{\mu}_{\mathrm{tar},<t}\right),
\end{aligned}
\end{equation}
where $\bar{\mu}_{\mathrm{tar}}=[\mu_{\mathrm{tar}},\circled{E}]$. The LLM is initialized from the CosyVoice2 checkpoint and trained on GigaEdit-S. We use gradient accumulation of 8, gradient clipping of 5, and a warmup learning-rate scheduler with 2,500 warmup steps. The learning rate is $1\times10^{-6}$, and the supervised LLM checkpoint used for GRPO is trained for 8 epochs.

\subsection{Flow Training}
We then adapt the Flow module to convert target speech tokens into Mel spectrograms while preserving the acoustic context of the original speech. Following CosyEdit~\cite{chen2026cosyedit}, we adopt Guided Optimal-Transport Conditional Flow Matching (GOT-CFM). Let $M_{\mathrm{ori}}$ and $M_{\mathrm{tar}}$ denote the Mel spectrograms of the original and target speech. We concatenate the original and target Mel states as
\begin{equation}
Z_0=[M_{\mathrm{ori}}^{0},M_{\mathrm{tar}}^{0}],
\qquad
Z_1=[M_{\mathrm{ori}},M_{\mathrm{tar}}],
\end{equation}
where $Z_0$ is sampled from the prior path and $Z_1$ is the data sample. The OT interpolation path and its target vector field are
\begin{align}
\phi_t^{\mathrm{OT}}(Z_0,Z_1) &= (1-t)Z_0+tZ_1, \\
\omega_t &= Z_1-Z_0 .
\end{align}

The Flow network predicts the vector field conditioned on the timestep $t$, speaker embedding $\mathbf{v}$, the up-sampled concatenated speech tokens $\mu_z=[\mu_{\mathrm{ori}},\mu_{\mathrm{tar}}]$, and the guided Mel condition $[M_{\mathrm{ori}},\tilde{M}_{\mathrm{tar}}]$:
\begin{equation}
\begin{aligned}
\nu_t
=
\mathrm{UNet}_{\theta}\bigl(
&\phi_t^{\mathrm{OT}}(Z_0,Z_1),\, t; \\
&\mathbf{v},\, \mu_z,\,
[M_{\mathrm{ori}},\tilde{M}_{\mathrm{tar}}]
\bigr),
\end{aligned}
\end{equation}
where $\tilde{M}_{\mathrm{tar}}$ is the masked target Mel spectrogram. The GOT-CFM objective minimizes the distance between the target and predicted vector fields:
\begin{equation}
\mathcal{L}_{\mathrm{GOT\text{-}CFM}}
=
\mathbb{E}_{t,Z_0,Z_1}
\left[
\left\|
\omega_t-\nu_t
\right\|_1
\right].
\end{equation}

Compared with ordinary target-speech generation, GOT-CFM exposes the Flow module to the complete original speech tokens and Mel spectrogram, providing a global acoustic guide for preserving leading and trailing silence, background noise, and other unedited acoustic contexts.

The Flow module is initialized from the CosyVoice2 checkpoint and also trained on GigaEdit-S. We use gradient accumulation of 8, gradient clipping of 5, and a constant learning rate of $3\times10^{-5}$. The Flow checkpoint used in CosyEdit2 is trained for 9 epochs.

\subsection{BigVGAN Training}
CosyVoice2 uses a HiFT-GAN vocoder with a 24~kHz waveform configuration, 80 Mel bands, hop size 480, and window/FFT size 1920. Since the official BigVGAN checkpoints do not provide a model with the same configuration, we cannot directly reuse a pretrained BigVGAN for CosyVoice2-style Mel spectrograms. To reduce the cost of training from scratch while preserving pretrained knowledge, we initialize BigVGAN from the closest available checkpoint, \texttt{bigvgan\_v2\_22khz\_80band\_256x}\footnote{\url{https://huggingface.co/nvidia/bigvgan_v2_22khz_80band_256x}}, and adapt its acoustic configuration to match CosyVoice2.

Specifically, we keep the 80-band Mel representation and modify the vocoder to operate at 24~kHz with hop size 480, window size 1920, FFT size 1920, and an overall upsampling ratio of 480. This allows BigVGAN to consume Mel spectrograms generated by the CosyEdit2 Flow module without changing the acoustic interface. Under this adaptation, we reuse 88.20\% of the generator parameters from the pretrained checkpoint, while the discriminator parameters are fully reused.

We train BigVGAN on a 625-hour mixed vocoder corpus. Specifically, it contains 585 hours from LibriTTS~\cite{panayotov2015librispeech} and LibriTTS-R~\cite{koizumi2023libritts}, including \texttt{train-clean-100} and \texttt{train-other-500} from LibriTTS and \texttt{train-clean-360} from LibriTTS-R, together with 40 hours randomly sampled from YODAS2~\cite{li2023yodas}. LibriTTS and LibriTTS-R provide high-quality multi-speaker read speech suitable for TTS vocoder training, while YODAS2 introduces long-form YouTube speech with more diverse in-the-wild acoustic conditions. This mixture exposes the vocoder to both clean speech and realistic background variations such as noise and music, which better matches speech editing scenarios where the generated waveform should preserve the recording characteristics of the original utterance. We train the adapted BigVGAN for 460k steps, substantially fewer than the 5M-step official pretraining, benefiting from the transferred pretrained parameters.

\begin{figure*}
    \centering
    \includegraphics[width=1\linewidth]{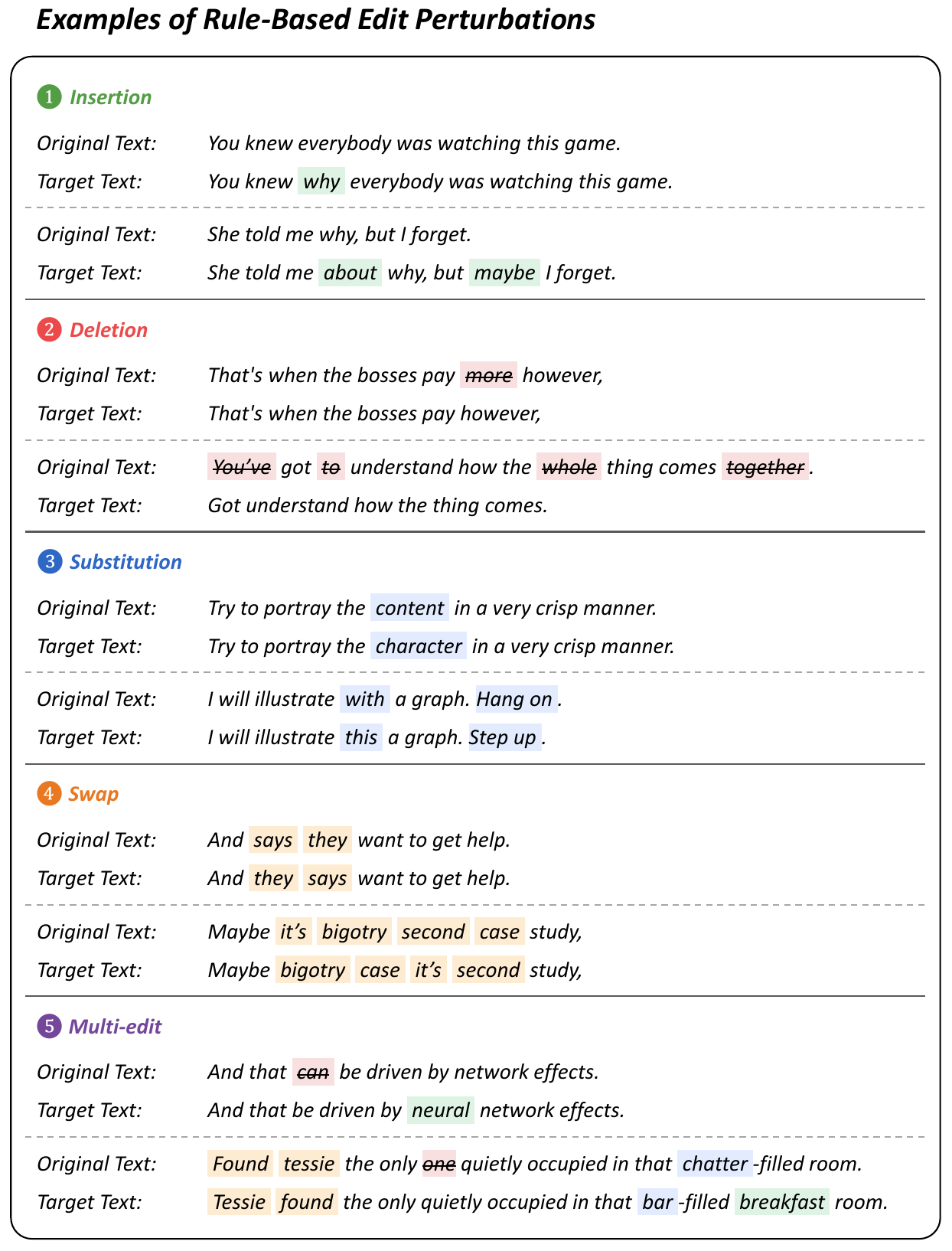}
    \caption{Examples of rule-based edit perturbations used in the TTS-to-edit prompt synthesis pipeline, including insertion, deletion, substitution, swap, and multi-edit operations.}
    \label{fig:edit_perturbations}
\end{figure*}

\begin{figure*}
    \centering
    \includegraphics[width=1\linewidth]{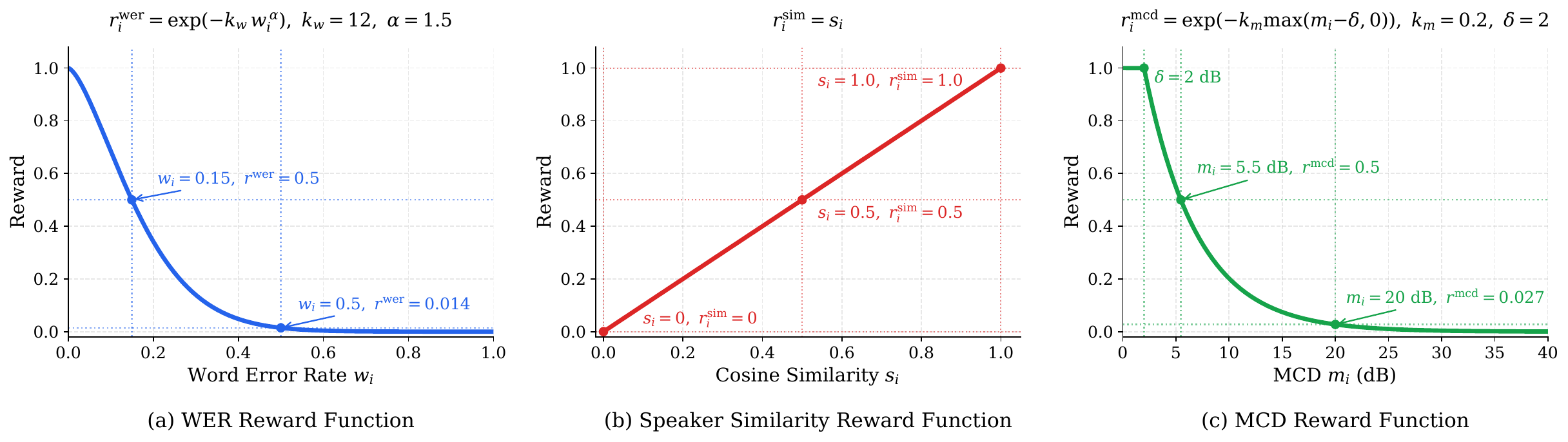}
    \caption{
        Reward functions used in the editing-oriented GRPO stage. 
        (a) The WER reward adopts an exponential decay with a power-law exponent, sharply penalizing high recognition errors while providing finer discrimination in the low-WER region. 
        (b) The speaker similarity reward directly uses cosine similarity as the reward value, preserving stable and interpretable ranking signals. 
        (c) The MCD reward introduces a tolerance margin $\delta$ before exponential decay, focusing optimization on preventing severe acoustic degradation in unedited regions rather than over-penalizing perceptually negligible variations.
    }
    \label{fig:reward_functions}
\end{figure*}

\section{Details for Stage 2}
\label{sec:appendix_data_construction_procedures}
\subsection{TTS-to-Edit Prompt Synthesis Pipeline}
The GRPO stage does not require paired target speech recordings. Instead, we automatically convert ordinary TTS-style speech-text pairs into speech editing prompts through rule-based textual perturbations. Specifically, given a speech waveform $Y_{\mathrm{ori}}$ and its transcription $X_{\mathrm{ori}}$, we synthesize a target text $X_{\mathrm{tar}}$ by applying text editing operations while keeping the original speech unchanged. The resulting triplet $(X_{\mathrm{ori}}, X_{\mathrm{tar}}, Y_{\mathrm{ori}})$ is then used as the editing prompt for GRPO training.

As illustrated in Figure~\ref{fig:edit_perturbations}, we implement five perturbation operations: insertion, deletion, substitution, swap, and multi-edit. The first four correspond to basic single-operation editing tasks, while multi-edit randomly composes multiple edits to construct more challenging editing instructions. All perturbations are implemented with rule-based NLP augmentation rather than LLM rewriting, enabling explicit control over edit type and edit length while avoiding semantic drift introduced by generative rewriting models.

For each utterance, an edit type is randomly sampled, and the edit length is adaptively determined according to the transcription length. Specifically, the maximum number of editable spans is constrained to at most half of the original word count:
\begin{equation}
N_{\mathrm{edit}} \le \max(1, \lfloor |X_{\mathrm{ori}}| / 2 \rfloor).
\end{equation}

For multi-edit augmentation, the number of edits is further randomly sampled to generate diverse combinations of insertion, deletion, substitution, and swap operations.

\subsection{Speech Token and Prompt Construction}

We adopt the speech-token extraction pipeline of CosyVoice2 to construct GRPO training prompts. Given the original speech waveform, we encode it into discrete speech tokens using the CosyVoice2 speech tokenizer. The editing condition is constructed as:
\[
[X_{\mathrm{ori}}, 
\; X_{\mathrm{tar}}, 
\; \mu_{\mathrm{ori}}],
\]
where the original and target texts are subsequently tokenized, sequentially concatenated, and then combined with the original speech-token sequence \(\mu_{\mathrm{ori}}\). During GRPO training, the LLM autoregressively predicts target speech tokens conditioned on this prompt, while the Flow and BigVGAN modules decode the generated tokens into waveforms for reward computation.

\input{tables/ming-freeform-audio-edit-en-full}

\subsection{ASR-Based Auxiliary Annotation}
To support reward computation and analysis, we additionally preprocess each utterance with WhisperX~\cite{bain2023whisperx}. Specifically, we first transcribe the original speech using large-v3-turbo~\cite{radford2023robust} and then perform forced alignment to obtain word-level timestamps, from which non-edited regions $\Omega$ are identified for MCD computation. The transcriptions are normalized using NeMo text normalization~\cite{zhang2021nemo}, contraction fixing, and punctuation removal for robustness. The resulting aligned ASR annotations are stored with the editing prompts for reward computation and error analysis.

Utterances with failed ASR alignment are automatically filtered out during preprocessing to ensure annotation consistency.

\subsection{Training Corpus}

For the GRPO stage, we sample 3{,}000 utterances from GigaSpeech-XL and synthesize editing prompts using the above pipeline. Since the construction process only requires ordinary speech-text pairs, the method is inherently scalable and can be applied to arbitrary TTS corpora without collecting manually edited target recordings. This target-speech-free design is particularly important for speech editing, where constructing perfectly matched target recordings with consistent speaker identity, prosody, and environmental acoustics is extremely difficult in practice.

\subsection{Reward Design Intuition}
\label{sec:appendix_reward_design}

Figure~\ref{fig:reward_functions} illustrates the reward functions used in the GRPO stage. Our reward design follows the practical preference hierarchy of speech editing: the generated speech should first satisfy the target editing instruction, then preserve the original acoustic characteristics in the unedited regions, and finally maintain speaker consistency.

The WER reward in Figure~\ref{fig:reward_functions}(a) adopts an exponential-decay formulation:
\begin{equation}
r_i^{\mathrm{wer}} = \exp(-k_w w_i^{\alpha}),
\end{equation}
which rapidly suppresses samples with large recognition errors. Compared with linear penalties, the exponential form provides stronger discrimination in the low-WER region, encouraging the policy to prioritize content correctness before optimizing finer acoustic properties. Empirically, we found that this sharp decay stabilizes early-stage GRPO training, where the generated speech may initially contain substantial recognition errors.

The speaker reward in Figure~\ref{fig:reward_functions}(b) directly uses cosine similarity as the reward value:
\begin{equation}
r_i^{\mathrm{sim}} = s_i.
\end{equation}

Unlike the WER and MCD rewards, speaker similarity already lies in a bounded and semantically meaningful range $[0,1]$, making additional nonlinear transformation unnecessary in our experiments. We therefore retain the original similarity score to preserve stable gradients and avoid over-amplifying noisy speaker variations.

MCD is computed via pymcd's DTW mode over non-edited regions $\Omega$. DTW alignment provides robustness against boundary imprecision introduced by forced alignment errors, yielding more reliable intra-group relative rankings for GRPO optimization. The resulting reward in Figure~\ref{fig:reward_functions}(c) introduces a tolerance margin $\delta$:
\begin{equation}
r_i^{\mathrm{mcd}} = \exp\!\left( -k_m \max(m_i-\delta,0) \right).
\end{equation}

This design reflects the observation that very small MCD differences are often perceptually negligible, while larger deviations usually correspond to noticeable prosody or timbre distortion in the non-edited regions. The thresholded exponential decay therefore focuses optimization on preventing severe acoustic degradation instead of over-penalizing minor variations.

Overall, the three rewards operate at different granularities: WER provides coarse-grained content supervision over the entire utterance, MCD constrains fine-grained acoustic preservation in the unchanged regions, and speaker similarity further ranks candidates according to global speaker consistency, particularly when multiple rollout candidates exhibit comparable performance on the former two metrics.

\input{tables/ming-freeform-audio-edit-zh}

\section{Additional Speech Editing Results}
\label{sec:appendix_speech_editing}
\subsection{Results on Ming-Freeform-Audio-Edit}
\label{sec:appendix_ming_results}
\paragraph{Baselines.}
We provide the full results on Ming-Freeform-Audio-Edit, extending the main-paper comparison with additional speech editing baselines. On the English subset, we include VoiceCraft, VoiceCraft-X, LEMAS-Edit, ECPA~\cite{ren2026edit}, SSR-Speech, Ming-UniAudio, and CosyEdit. On the Chinese subset, we compare with multilingual or Chinese-capable systems, including VoiceCraft-X, LEMAS-Edit, and Ming-UniAudio. All systems are evaluated with their recommended inference settings when available. For ECPA, the model is not publicly released, so we report the results from the original paper.

\paragraph{Full Results on English Subset.}

Table~\ref{tab:editing_results_en_dnsmos_mae_split_full} reports the complete English results. The overall trend is consistent with the main paper: CosyEdit2 consistently outperforms multilingual systems such as VoiceCraft-X, LEMAS-Edit, and Ming-UniAudio, and remains competitive with the strong monolingual cascaded system SSR-Speech. CosyEdit2 performs particularly well on substitution, achieving the best WER on both splits while matching the best speaker similarity. For insertion, it approaches SSR-Speech in WER and SS, while clearly obtaining lower $\mathrm{MAE}_{\mathrm{DNSMOS}}$. For deletion, SSR-Speech remains slightly stronger in WER and SS, likely benefiting from explicit speech-text alignment for locating deleted content, but CosyEdit2 still yields the best acoustic-quality consistency.

Compared with ECPA, which also explores GRPO for speech editing, CosyEdit2 demonstrates superior content accuracy and speaker preservation due to a fundamental difference in optimization. While ECPA relies on a pretrained TTS model as an implicit critic to optimize semantic-prosodic self-consistency under a general synthesis prior, CosyEdit2 utilizes an editing-oriented GRPO framework for teacher-free, outcome-level optimization. By directly rewarding semantic correctness and acoustic preservation on decoded speech, our approach explicitly targets speech-editing preferences rather than general generation distributions. Consequently, CosyEdit2 effectively mitigates editing artifacts and contextual mismatch, yielding lower WER and higher speaker similarity across all edit types while avoiding the acoustic normalization risks inherent in generic TTS priors.

Absolute DNSMOS scores can be deceptive in text-based speech editing, as elevated metrics often reflect acoustic normalization or global denoising rather than faithful contextual preservation. Such elevated scores are more reflective of output cleanliness and acoustic normalization than of how well the edited segment blends into its surrounding context, as global denoising naturally inflates perceptual quality metrics regardless of local consistency. However, imperceptible editing requires modified segments to seamlessly match the acoustic environment and stylistic texture of the unedited context. By directly optimizing for editing-specific rewards, CosyEdit2 maintains competitive speech quality while ensuring strict contextual alignment. This distinction underscores that absolute perceptual scores cannot independently verify editing fidelity, highlighting the necessity of consistency-focused metrics to gauge non-disruptive boundary fusion.

\paragraph{Results on Chinese Subset.}
\input{tables/realedit}

Table~\ref{tab:editing_results_zh_dnsmos_mae_split} reports the Chinese results. CosyEdit2 substantially outperforms all multilingual baselines on WER and SS across insertion, deletion, and substitution, demonstrating strong cross-lingual speech editing capability. The gains are especially large for insertion and substitution, where CosyEdit2 reduces WER to around 1--1.4\% while maintaining the highest speaker similarity. Deletion remains more difficult, but CosyEdit2 still achieves the best WER and SS, showing that the model generalizes beyond English despite being optimized with English GRPO prompts. In terms of acoustic consistency, CosyEdit2 obtains the lowest $\mathrm{MAE}_{\mathrm{DNSMOS}}$ on insertion and substitution and remains competitive on deletion. These results indicate that the editing-oriented training improves not only content modification but also speaker and acoustic preservation in multilingual speech editing.

\subsection{Results on RealEdit}
\label{sec:appendix_realedit_results}

\paragraph{Baselines.}
We further evaluate CosyEdit2 on the RealEdit benchmark from VoiceCraft~\cite{peng2024voicecraft}. We compare with three cascaded editing systems, including FluentSpeech, VoiceCraft, and SSR-Speech, as well as two end-to-end speech editing models, Ming-UniAudio and CosyEdit. All baseline results are obtained under their recommended settings.

\paragraph{Evaluation Setup and Metrics.}
RealEdit contains 310 in-the-wild speech editing cases with realistic acoustic variations such as background noise and music. We report WER for content accuracy, speaker similarity (SS) for speaker preservation, and MCD on the unedited regions for acoustic preservation. Following prior RealEdit evaluation, we use MOSNet~\cite{lo2019mosnet} to estimate perceptual quality and additionally report its mean absolute error against the ground-truth speech, where lower MAE indicates closer quality consistency.

\paragraph{Results.}
Table~\ref{tab:realedit_results} reports the RealEdit results. CosyEdit2 achieves the lowest WER among all systems, improving over both strong cascaded editors such as SSR-Speech and end-to-end models such as Ming-UniAudio and CosyEdit. It also obtains the best speaker similarity among end-to-end systems and approaches the strongest cascaded baseline.

More importantly, CosyEdit2 substantially reduces MCD on the unedited regions, from 4.94 to 3.93 compared with CosyEdit. This confirms that the proposed editing-oriented training improves acoustic preservation rather than merely optimizing text correctness. In terms of MOSNet, CosyEdit2 achieves a low $\mathrm{MAE}_{\mathrm{MOSNet}}$ to the ground truth, indicating that it better matches the original recording quality under in-the-wild conditions. Overall, the RealEdit results further show that CosyEdit2 combines the content accuracy of strong editing systems with stronger acoustic preservation in realistic speech editing scenarios.

\input{tables/seed-tts-eval}

\section{Additional Zero-Shot TTS Results on SEED-TTS-EVAL}
\label{sec:appendix_seed_tts_eval}

\subsection{Evaluation Setup and Metrics.}
We further evaluate zero-shot TTS on SEED-TTS-EVAL~\cite{anastassiou2024seed} benchmark, which contains English and Chinese test sets for measuring content intelligibility and speaker similarity. We use the same inference setting as in the main zero-shot TTS experiments: CosyEdit2 replaces only the LLM with the GRPO-optimized one, while keeping the original CosyVoice2 Flow and HiFT-GAN vocoder unchanged. Unlike CV3-EVAL, SEED-TTS-EVAL prompts are generally clean and well-trimmed, so no VAD-based preprocessing is applied in this evaluation. Following the SEED-TTS-EVAL protocol, we report CER (\%) for Chinese, WER (\%) for English, and speaker similarity (SS, \%) for voice cloning fidelity.

\begin{figure*}
    \centering
    \includegraphics[width=1\linewidth]{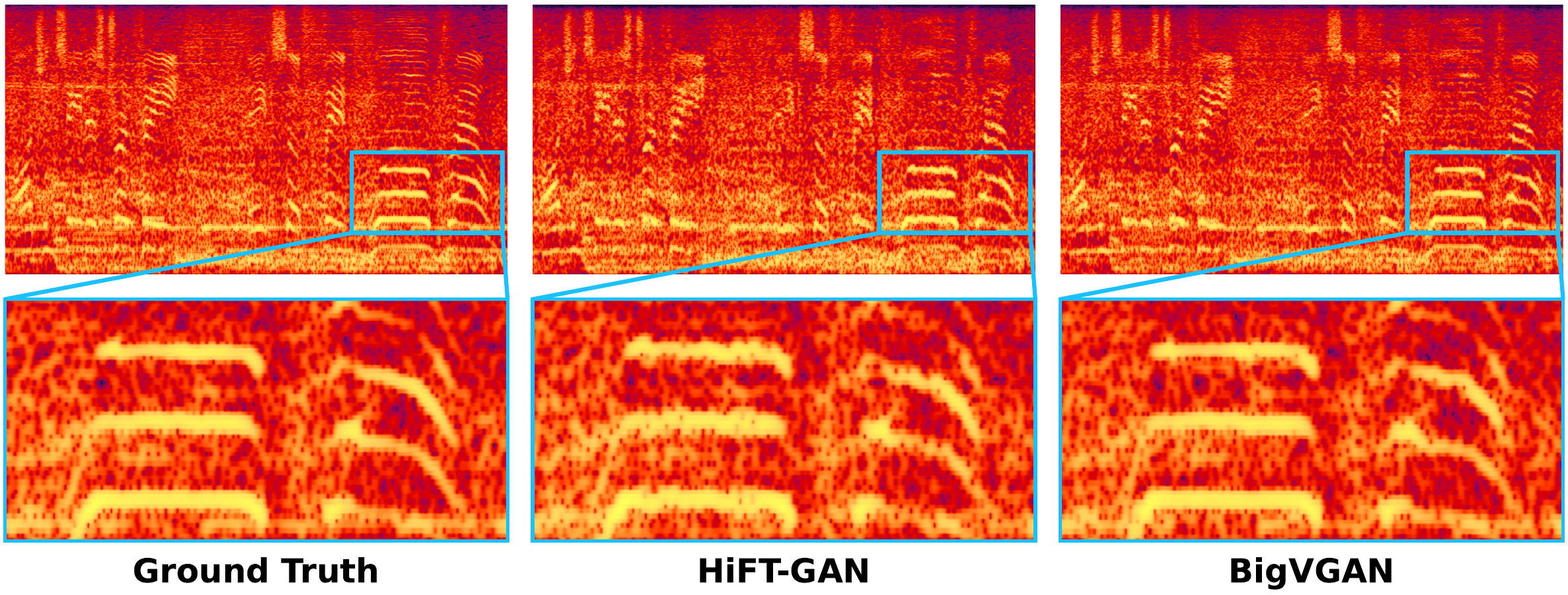}
    \caption{Mel spectrogram visualization of a speech sample reconstructed by HiFT-GAN from CosyVoice2 and our trained BigVGAN, with a zoomed-in view of harmonic components.}
    \label{fig:vocoder_comparison}
\end{figure*}

\input{tables/vocoder_reconstruction}

\subsection{Results.}
Table~\ref{tab:seed_eval_results} shows the results on SEED-TTS-EVAL benchmark. CosyEdit2 achieves the best content accuracy in both languages, reducing CER from 1.36 to 1.16 on test-zh and WER from 3.10 to 1.95 on test-en compared with the same-backbone CosyVoice2 baseline. Speaker similarity is also well preserved: CosyEdit2 obtains the highest SS on test-zh and remains competitive on test-en.

These results further support our finding that editing-oriented GRPO strengthens the zero-shot generation ability of the backbone, even though it is trained with editing prompts rather than zero-shot TTS supervision. Notably, the improvement transfers from English GRPO training to Chinese zero-shot TTS, suggesting that the learned gains are not language-specific. Instead, GRPO appears to enhance general speech generation abilities, such as speech-text alignment and pronunciation control, which benefit zero-shot TTS across languages without substantially degrading voice cloning fidelity.

\section{Vocoder Reconstruction Experiment}
\label{sec:vocoder_reconstruction}
To isolate the effect of the vocoder, we conduct a reconstruction experiment comparing our adapted BigVGAN with the original HiFT-GAN vocoder from CosyVoice2. Given a source waveform, we first extract its Mel spectrogram with the corresponding vocoder frontend and then reconstruct the waveform using the vocoder. The reconstructed audio is compared against the same source waveform, this experiment evaluates Mel-to-waveform reconstruction quality without involving the LLM or Flow modules.

\subsection{Evaluation Setup}
The test split of VoiceBank-DEMAND-16k\footnote{\url{https://huggingface.co/datasets/JacobLinCool/VoiceBank-DEMAND-16k}} serves as our evaluation benchmark, covering both clean and noisy subsets as source audio. For each subset, the source waveform is used to compute the Mel input and also serves as the reconstruction reference. Since the dataset is 16~kHz while both vocoders operate at the CosyVoice2 acoustic configuration of 24~kHz, we resample the source audio from 16~kHz to 24~kHz for Mel extraction and waveform generation, and then resample the generated waveform back to 16~kHz for evaluation. All generated waveforms are length-matched with the reference audio before metric computation.

\subsection{Metrics}
We report reference-based reconstruction metrics, including multi-resolution short-time Fourier transform distance (MR-STFT)~\cite{yamamoto2020parallel}, PESQ~\cite{rix2001perceptual}, STOI~\cite{taal2011algorithm}, ESTOI~\cite{jensen2016algorithm}, and MCD~\cite{kubichek1993mel}. MR-STFT and MCD measure spectral distortion, PESQ estimates perceptual speech quality, and STOI/ESTOI measure intelligibility preservation. In addition, We compute DNSMOS on both the source and reconstructed audio, and report $\mathrm{MAE}_{\mathrm{DNSMOS}}$ to measure how closely the reconstructed waveform preserves the source perceptual-quality score.

\subsection{Results}
Table~\ref{tab:vocoder_reconstruction} shows that BigVGAN consistently outperforms the original HiFT-GAN vocoder on both clean and noisy subsets. On clean speech, BigVGAN improves all reference-based metrics, reducing MCD from 1.631 to 1.310 and MR-STFT from 1.215 to 1.138, while increasing PESQ, STOI, and ESTOI. On noisy speech, BigVGAN also yields better reconstruction quality, reducing MCD from 1.988 to 1.630 and improving PESQ from 3.019 to 3.185. These gains indicate that the adapted BigVGAN reconstructs Mel spectrograms more faithfully than the original HiFT-GAN.

The improvement is especially important for speech editing. In editing scenarios, the vocoder should not merely synthesize clean speech, but should preserve the acoustic characteristics contained in the generated Mel spectrogram, including background noise and other in-the-wild recording conditions. BigVGAN improves all metrics on both clean and noisy subsets, with especially clear reductions in MCD, indicating more faithful spectral reconstruction and stronger acoustic preservation. The lower $\mathrm{MAE}_{\mathrm{DNSMOS}}$ suggests better perceptual-quality consistency with the source audio. This supports our replacement of HiFT-GAN with BigVGAN in CosyEdit2.

We also provide a visual comparison in Figure~\ref{fig:vocoder_comparison}, showing the Mel spectrograms reconstructed by HiFT-GAN vocoder from CosyVoice2 and our trained BigVGAN on a challenging sample that contains simultaneous speech, background music, and ambient noise. In the zoomed-in region, HiFT-GAN produces noticeably blurred harmonic structures, with individual harmonics appearing smeared, less well-defined, and in some cases entirely absent. In contrast, BigVGAN reconstructs sharper and more clearly delineated harmonic components, preserving structures that HiFT-GAN fails to reproduce, and more closely resembling the ground truth. This qualitative observation is consistent with the quantitative improvements in MCD and MR-STFT reported above.

\section{Speech Preservation Evaluation}
\label{sec:speech_preservation_evaluation}
To further examine the preservation ability of different systems, we design a text-identity reconstruction experiment, where the target text is identical to the original or prompt text. This setting removes the need for content modification and directly evaluates whether a model can reproduce the original speech. Notably, our training data does not contain such identity-editing samples, making this experiment a zero-shot test of acoustic preservation.

\subsection{Evaluation Setup}
We evaluate on RealEdit, which contains 310 in-the-wild speech samples with complex acoustic conditions. For CosyEdit2, we set the original text and target text to be identical, with $X_{\mathrm{tar}}=X_{\mathrm{ori}}$, and generate the target speech using the speech editing pipeline. For CosyVoice2, we use its zero-shot TTS mode with the prompt text and target text set to the same transcription. Thus, both models are asked to reconstruct the original utterance content, but with different task formulations: CosyVoice2 regenerates the utterance as zero-shot TTS, while CosyEdit2 treats it as an identity edit.

We also include two oracle vocoder reconstruction upper bounds. For HiFT-GAN and BigVGAN, we extract the Mel spectrogram from the original speech and reconstruct the waveform with the corresponding vocoder. These oracle settings bypass the LLM and Flow modules, thus reflect the reconstruction upper bound of the acoustic backend.

\subsection{Metrics}

We report speaker similarity (SS) and Mel-Cepstral Distortion (MCD) between the generated speech and the original speech. SS measures whether the reconstructed speech preserves the speaker identity, while MCD measures spectral distortion over the full utterance. Higher SS and lower MCD indicate stronger preservation.

\input{tables/reconstruction-fidelity-experiment}

\input{tables/ming-en-subjective-evaluation}
\input{tables/ming-zh-subjective-evaluation}

\subsection{Results}

Table~\ref{tab:reconstruction-fidelity-experiment} shows the preservation evaluation results. CosyEdit2 substantially improves over the same-backbone CosyVoice2 baseline, increasing SS from 96.92 to 99.08 and reducing MCD from 6.24 to 3.07. This confirms that the editing-oriented training greatly strengthens the model's ability to preserve the original speech, even without explicit identity-editing examples during training. In contrast, zero-shot TTS can reproduce the content and speaker identity, but tends to regenerate the utterance with weaker preservation of the original acoustic trajectory.

Compared with the oracle vocoder reconstructions, CosyEdit2 is already close to the HiFT-GAN upper bound and only slightly behind the BigVGAN upper bound in MCD. Notably, CosyEdit2 even slightly surpasses the HiFT-GAN reconstruction in speaker similarity, achieving 99.08 versus 99.02. Since the oracle BigVGAN reconstruction directly uses the Mel spectrogram extracted from the original speech, the remaining gap mainly reflects errors from the LLM and Flow modules. The small difference between CosyEdit2 and the oracle reconstructions indicates that the LLM and Flow preserve acoustic information effectively under the identity-editing setting.

These results provide additional evidence for our central claim: speech editing requires stronger preservation than ordinary zero-shot TTS. When no content change is needed, CosyEdit2 behaves close to an acoustic reconstruction system, while CosyVoice2 still behaves like a generative TTS model. This supports our view that editing-oriented training improves not only content modification, but also the preservation ability required to avoid degeneration into ordinary zero-shot TTS.

\section{Subjective Evaluation}
\label{sec:appendix_subjective_evaluation}
\subsection{Annotators}
\begin{figure*}
    \centering
    \includegraphics[width=1\linewidth]{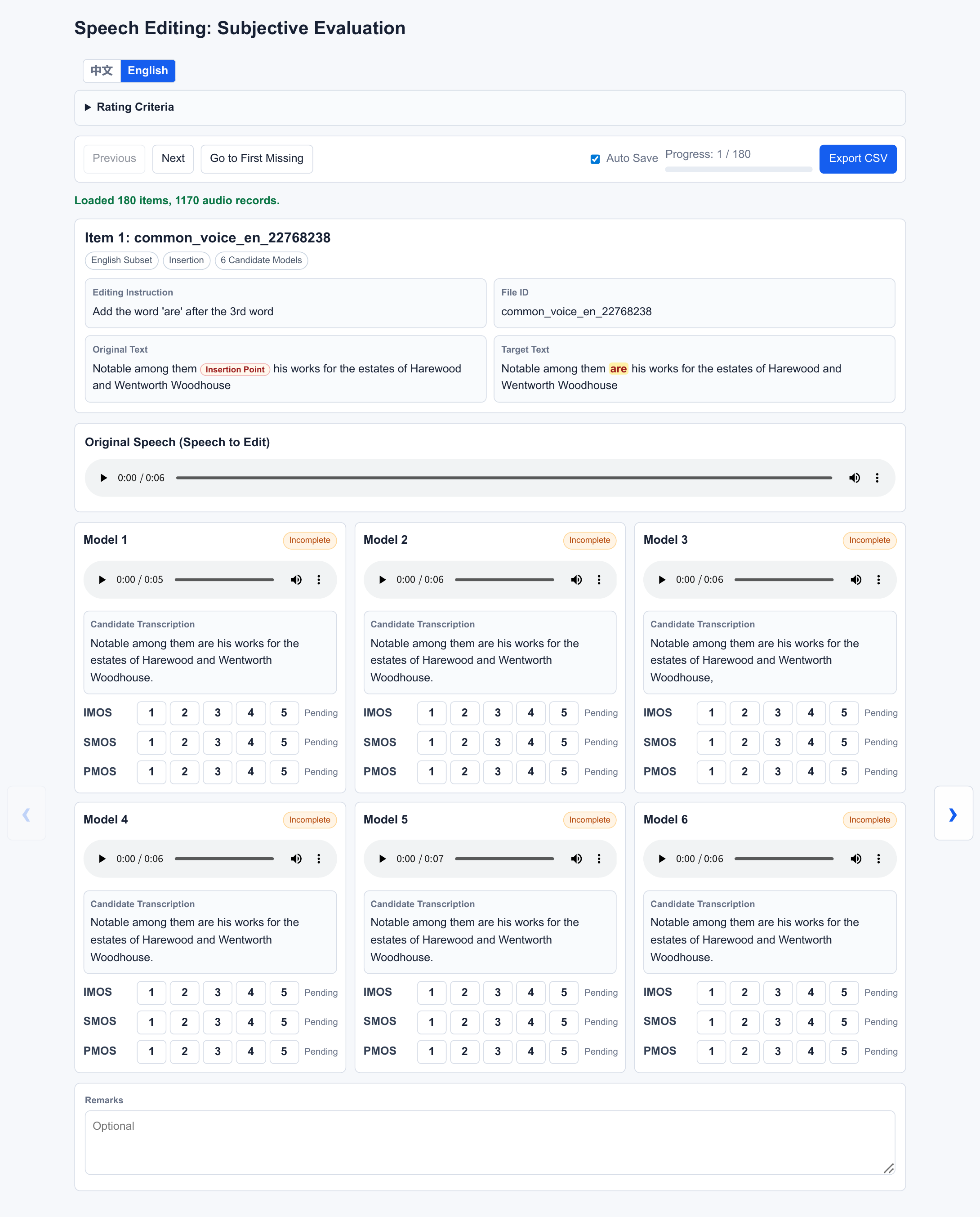}
    \caption{Speech Editing Subjective Evaluation Annotation UI.}
    \label{fig:subjective_evaluation-speech-editing}
\end{figure*}

\begin{figure*}
    \centering
    \includegraphics[width=1\linewidth]{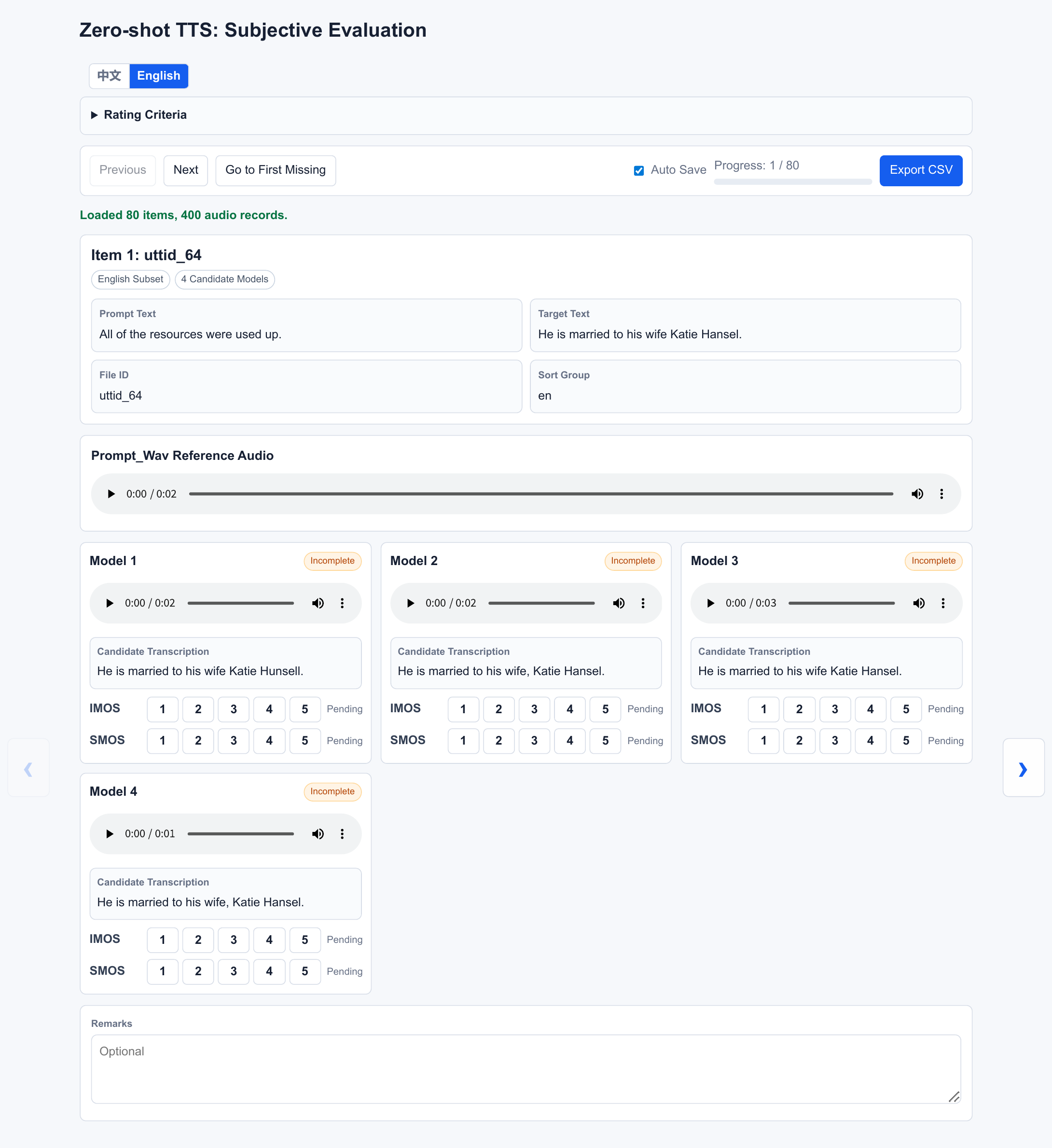}
    \caption{Zero-shot TTS Subjective Evaluation Annotation UI.}
    \label{fig:subjective_evaluation-zero-shot-tts}
\end{figure*}

We manually recruited 10 university students for subjective evaluation, including 5 male and 5 female annotators. All annotators are native Chinese speakers and have passed CET-6 or an equivalent English proficiency level. Each audio sample is rated independently, and different metrics are judged separately without compensating one dimension by another. The annotation interfaces for speech editing and zero-shot TTS are shown in Figure~\ref{fig:subjective_evaluation-speech-editing} and Figure~\ref{fig:subjective_evaluation-zero-shot-tts}, respectively.

\subsection{Rating Criteria}
\paragraph{Zero-shot TTS.}
For zero-shot TTS, annotators rate each generated sample using two Mean Opinion Score (MOS) metrics on a 1--5 integer scale: Intelligibility MOS (IMOS) and Speaker Similarity MOS (SMOS). IMOS measures content intelligibility and consistency with the target text, while SMOS measures speaker similarity to the prompt speech. The prompt speech is provided as the reference speaker, and candidate transcriptions are shown only to help identify possible content errors; final judgments are based on the target text and generated target speech.

For IMOS, the scores are defined as follows:
\begin{itemize}
    \item 5: The speech is clear and natural, and the content fully matches the target text without omissions, insertions, substitutions, repetitions, or semantic errors.
    \item 4: The speech is mostly clear and faithful to the target text, with only minor pronunciation, pause, or content deviations that do not affect understanding.
    \item 3: The speech is generally intelligible, but contains noticeable content mismatches such as omissions, substitutions, repetitions, or ambiguous pronunciations.
    \item 2: The speech is difficult to understand and has clear mismatches with the target text, including multiple missing, incorrect, repeated, or broken segments.
    \item 1: Most content is unintelligible or severely deviates from the target text.
\end{itemize}
For SMOS, the scores are defined as follows:
\begin{itemize}
    \item 5: The generated speech is almost identical to the prompt speaker in identity, timbre, pitch, speaking style, and vocal manner.
    \item 4: The speaker similarity is high, with only slight differences in timbre, pitch, or speaking style.
    \item 3: The speaker identity is partly preserved, but noticeable differences remain.
    \item 2: The generated speech has low speaker similarity and clearly deviates from the prompt speaker.
    \item 1: The generated speech sounds like a different speaker.
\end{itemize}

\paragraph{Speech Editing.}
For speech editing, annotators rate each edited sample with IMOS, SMOS, and PMOS (Preservation MOS). IMOS measures whether the edited speech matches the target text, SMOS measures speaker preservation with respect to the original speech, and PMOS measures preservation of unedited regions and edit-boundary naturalness.

The IMOS and SMOS scales follow the same principles as in zero-shot TTS, except that the reference is the original speech to be edited. PMOS is defined as follows:
\begin{itemize}
    \item 5: The unedited regions are almost identical to the original speech in timbre, prosody, speaking rate, volume, background condition, and recording style; edit boundaries are natural and inaudible.
    \item 4: The unedited regions are well preserved, with only slight differences in timbre, prosody, acoustic condition, or boundary smoothness.
    \item 3: Noticeable changes exist in the unedited regions, such as altered prosody, speaking rate, volume, background condition, or mildly discontinuous edit boundaries.
    \item 2: The unedited regions are poorly preserved, or the edit boundaries contain clear artifacts, abrupt changes, or discontinuities.
    \item 1: The unedited regions are largely resynthesized or severely deviate from the original speech, or the edit boundaries are highly unnatural.
\end{itemize}

\subsection{Speech Editing Subjective Evaluation}

\paragraph{Task Design.}
Subjective evaluations were conducted on the English and Chinese subsets of Ming-Freeform-Audio-Edit. For each language, 30 utterances were randomly sampled for each edit type (insertion, deletion, and substitution), resulting in 90 samples per language and 180 samples in total. The evaluated systems were identical to those in the objective speech editing experiments. Additionally, CosyVoice2 in its zero-shot TTS mode was included as a same-backbone baseline.

\paragraph{Results.}
Tables~\ref{tab:subjective_results_en} and~\ref{tab:subjective_results_zh} show the subjective speech editing results. On the English subset, CosyEdit2 achieves the best IMOS and SMOS across all edit types, and obtains the best or nearly best PMOS. It is especially strong on substitution, where it achieves the highest scores on all three dimensions. Compared with SSR-Speech, CosyEdit2 reaches comparable preservation quality in the reconstructed unedited regions, demonstrating that, from the perspective of human auditory perception, our end-to-end speech editing model can achieve a similar level of preservation fidelity. Compared with CosyVoice2, CosyEdit2 consistently improves PMOS, showing that ordinary zero-shot TTS can generate intelligible edited content but fails to preserve unedited regions as reliably.

On the Chinese subset, CosyEdit2 achieves the best IMOS, SMOS, and PMOS for all edit types. The improvement over CosyVoice2 is particularly clear in PMOS, confirming that the editing-oriented training strengthens preservation of non-edited regions rather than merely improving content generation. These subjective results support the main conclusion of the objective experiments: CosyEdit2 improves speech editing quality not only by producing correct target content, but also by better preserving speaker identity, acoustic conditions, and edit-boundary naturalness.

\subsection{Zero-shot TTS Subjective Evaluation}
\input{tables/cv3-eval-subjective-evaluation}
\paragraph{Task Design.}
For zero-shot TTS, we conducted a subjective evaluation on the multilingual voice cloning subset of CV3-EVAL. We randomly sampled 20 utterances from each of the \texttt{zh}, \texttt{en}, \texttt{hard\_zh}, and \texttt{hard\_en} subsets, resulting in 80 samples in total. The evaluated systems were selected from the objective zero-shot TTS experiments. For reporting purposes, the \texttt{zh} and \texttt{hard\_zh} subsets were aggregated into Chinese, while the \texttt{en} and \texttt{hard\_en} subsets were aggregated into English.

\paragraph{Results.}
Table~\ref{tab:zeroshot_tts_subjective} reports the subjective zero-shot TTS results. CosyEdit2 achieves the highest IMOS and SMOS in both English and Chinese, outperforming the same-backbone CosyVoice2 baseline. This indicates that editing-oriented GRPO improves perceived content correctness while maintaining or improving speaker similarity. The gains are consistent with the objective results, further showing that training with editing prompts does not degrade zero-shot TTS perceptual quality; instead, it improves the shared in-context learning capability underlying prompt-conditioned speech generation in a way that transfers to both English and Chinese synthesis.

\end{document}

%% file: tables/ming-freeform-audio-edit-en.tex
\begin{table*}[t]
\centering
\small
\renewcommand{\arraystretch}{0.98}
\resizebox{\textwidth}{!}{
\begin{tabular}{llcccc}
\toprule
Edit Type & Model & \multicolumn{4}{c}{Performance} \\
\midrule
\perfhead
\multirow{5}{*}{Insertion}
& GroundTruth             & \pair{--}{--}       & \pair{--}{--}       & \pair{2.99}{3.01} & \pair{--}{--} \\
& VoiceCraft-X   & \pair{5.74}{6.27} & \pair{0.85}{0.84} & \pair{3.05}{3.06} & \pair{0.166}{0.162} \\
& SSR-Speech     & \pair{\textbf{1.75}}{2.03} & \pair{\textbf{0.94}}{\textbf{0.94}} & \pair{3.06}{3.06} & \pair{0.139}{0.128} \\
& Ming-UniAudio  & \pair{6.49}{7.84} & \pair{0.80}{0.80} & \pair{3.04}{3.04} & \pair{0.168}{0.167} \\
& CosyEdit2      & \pair{1.90}{\textbf{1.93}} & \pair{0.93}{0.93} & \pair{3.02}{3.03} & \pair{\textbf{0.107}}{\textbf{0.108}} \\
\midrule

\perfhead
\multirow{5}{*}{Deletion}
& GroundTruth             & \pair{--}{--}       & \pair{--}{--}       & \pair{3.05}{3.03} & \pair{--}{--} \\
& VoiceCraft-X   & \pair{9.71}{10.65} & \pair{0.80}{0.80} & \pair{3.01}{3.00} & \pair{0.172}{0.173} \\
& SSR-Speech     & \pair{\textbf{5.22}}{\textbf{5.29}} & \pair{\textbf{0.91}}{\textbf{0.91}} & \pair{3.03}{3.02} & \pair{0.132}{0.134} \\
& Ming-UniAudio  & \pair{14.79}{24.37} & \pair{0.77}{0.75} & \pair{2.97}{2.97} & \pair{0.206}{0.204} \\
& CosyEdit2      & \pair{5.52}{5.83} & \pair{0.90}{0.90} & \pair{3.01}{3.00} & \pair{\textbf{0.131}}{\textbf{0.131}} \\
\midrule

\perfhead
\multirow{5}{*}{Substitution}
& GroundTruth             & \pair{--}{--}       & \pair{--}{--}       & \pair{3.04}{3.05} & \pair{--}{--} \\
& VoiceCraft-X   & \pair{7.29}{6.73} & \pair{0.81}{0.81} & \pair{3.04}{3.05} & \pair{0.155}{0.146} \\
& SSR-Speech     & \pair{1.90}{1.95} & \pair{\textbf{0.89}}{\textbf{0.90}} & \pair{3.08}{3.08} & \pair{0.146}{0.140} \\
& Ming-UniAudio  & \pair{8.10}{7.95} & \pair{0.77}{0.77} & \pair{3.00}{3.03} & \pair{0.166}{0.178} \\
& CosyEdit2      & \pair{\textbf{1.43}}{\textbf{1.52}} & \pair{\textbf{0.89}}{\textbf{0.90}} & \pair{3.04}{3.05} & \pair{\textbf{0.137}}{\textbf{0.132}} \\
\bottomrule
\end{tabular}
}
\caption{Performance comparison on the English subset of Ming-Freeform-Audio-Edit. 
MAE denotes $\mathrm{MAE}_{\mathrm{DNSMOS}}$ between generated and original speech, where lower is better.}
\label{tab:editing_results_en_dnsmos_mae_split}
\end{table*}

%% file: tables/ablation-experiment.tex
\begin{table*}[t]
\centering
\small
\renewcommand{\arraystretch}{0.98}
\begin{tabular}{l ccc ccccc}
\toprule
\multirow{2}{*}{Method} 
    & \multicolumn{3}{c}{Components} 
    & \multirow{2}{*}{WER $\downarrow$} 
    & \multirow{2}{*}{SS $\uparrow$} 
    & \multirow{2}{*}{MCD $\downarrow$} 
    & \multirow{2}{*}{DNSMOS} 
    & \multirow{2}{*}{MAE $\downarrow$} \\
\cmidrule(lr){2-4}
& LLM & Flow & BigVGAN & & & & & \\
\midrule
GT
    & -- & -- & --
    & 6.06 & -- & -- & 3.03 & -- \\
\midrule
CosyVoice2
    & $\times$ & $\times$ & $\times$
    & \textbf{4.14} & 96.65 & 6.68 & 3.28 & 0.275 \\
\midrule
\multirow{5}{*}{CosyEdit2}
    & SFT  & $\times$   & $\times$
    & 5.83 & 97.05 & 5.82 & 3.20 & 0.207 \\
    & GRPO & $\times$   & $\times$
    & 4.71 & 97.23 & 5.50 & 3.20 & 0.210 \\
    & GRPO & \checkmark & $\times$
    & 4.34 & 97.79 & 4.07 & 3.07 & 0.134 \\
    & GRPO & $\times$   & \checkmark
    & 4.69 & 97.27 & 5.42 & 3.21 & 0.208 \\
    & GRPO & \checkmark & \checkmark
    & 4.31 & \textbf{97.91} & \textbf{3.93} & 3.04 & \textbf{0.131} \\
\bottomrule
\end{tabular}
\caption{Ablation results for CosyEdit2 on RealEdit. MAE denotes $\mathrm{MAE}_{\mathrm{DNSMOS}}$ between generated and original speech.
    \textbf{LLM}: training strategy of the language model (GRPO or SFT).
    \textbf{Flow}: \checkmark\ indicates the flow matching module is fine-tuned for speech editing; 
    $\times$ indicates it uses the original pre-trained weights.
    \textbf{BigVGAN}: \checkmark\ indicates BigVGAN is used as the vocoder; 
    $\times$ indicates falling back to the HiFtGAN vocoder from CosyVoice2.}
\label{tab:ablation_realedit}
\end{table*}

%% file: tables/cv3-eval-simple.tex
\begin{table}[t]
\centering
\small
\renewcommand{\arraystretch}{0.98}
\begin{tabular}{lcccc}
\toprule
Model & zh & en & ja & ko \\
\midrule
VoiceCraft-X & 9.74  & 19.17 & 23.00 & 43.15 \\
SSR-Speech   & --    & 14.08 & --    & --    \\
CosyEdit     & 5.29  & 7.22  & 17.31 & 9.25  \\
CosyVoice2   & 3.77  & 5.45  & 7.76  & 6.89  \\
\midrule
CosyEdit2    & \textbf{3.52} & \textbf{4.87} & \textbf{6.16} & \textbf{5.14} \\
\bottomrule
\end{tabular}
\caption{CER(\%) and WER(\%) on the CV3-Eval Multilingual Voice Cloning subset.}
\label{tab:cv3_eval_multilingual_vc}
\end{table}

%% file: tables/cv3-eval-hard.tex
\begin{table*}[t]
\centering
\small
\renewcommand{\arraystretch}{0.98}
\begin{tabular}{lcccccc}
\toprule
\multirow{2}{*}{Model}
& \multicolumn{3}{c}{hard-zh}
& \multicolumn{3}{c}{hard-en} \\
\cmidrule(lr){2-4} \cmidrule(lr){5-7}
& CER (\%) $\downarrow$ & SS $\uparrow$ & DNSMOS $\uparrow$
& WER (\%) $\downarrow$ & SS $\uparrow$ & DNSMOS $\uparrow$ \\
\midrule
VoiceCraft-X & 26.36 & 65.74 & 3.74 & 28.64 & 55.68 & 3.75 \\
SSR-Speech   & --    & --    & --   & 28.96 & 42.30 & 3.70 \\
CosyEdit     & 34.60 & 74.13 & 3.52 & 13.93 & 67.16 & 3.88 \\
CosyVoice2   & 15.70 & 74.96 & 3.84 & 8.11  & 67.22 & \textbf{3.97} \\
\midrule
CosyEdit2    & \textbf{8.06} & \textbf{75.86} & \textbf{3.85}
             & \textbf{5.93} & \textbf{67.32} & \textbf{3.97} \\
\ \ \ \ - w/o GRPO    & 13.17 & 75.17 & 3.80
             & 11.99 & 67.24 & 3.93 \\

\bottomrule
\end{tabular}
\caption{CER(\%), WER(\%), Speaker Similarity (SS, \%), and DNSMOS scores on the hard samples in the CV3-Eval Multilingual Voice Cloning subset. w/o GRPO corresponds to the Stage 1 SFT model.}
\label{tab:cv3_eval_hard_multilingual_vc}
\end{table*}

%% file: tables/cv3-eval-coss-lingual.tex
\begin{table*}[t]
\centering
\small
\renewcommand{\arraystretch}{0.98}
\begin{tabular}{lcccccccccc}
\toprule
\multirow{2}{*}{Model}
& \multicolumn{2}{c}{to-zh}
& \multicolumn{3}{c}{to-en}
& \multicolumn{2}{c}{to-ja}
& \multicolumn{3}{c}{to-ko} \\
\cmidrule(lr){2-3}
\cmidrule(lr){4-6}
\cmidrule(lr){7-8}
\cmidrule(lr){9-11}
& en & ko
& zh & ja & ko
& en & ko
& zh & en & ja \\
\midrule
VoiceCraft-X
& 39.66 & 44.06
& 23.12 & 49.87 & 41.77
& 27.98 & 24.25
& 36.51 & 54.02 & 51.71 \\

CosyEdit
& 59.78 & 17.86
& 10.74 & 16.49 & 10.38
& 30.04 & 32.34
& 30.60 & 34.23 & 22.82 \\

CosyVoice2
& 13.15 & 6.51
& 6.02 & 13.45 & 7.92
& 15.13 & 7.85
& 8.35 & 10.40 & 10.94 \\

\midrule

CosyEdit2
& \textbf{7.16} & \textbf{4.43}
& \textbf{5.07} & \textbf{7.22} & \textbf{5.68}
& \textbf{14.12} & \textbf{5.75}
& \textbf{5.91} & \textbf{6.00} & \textbf{8.40} \\
\bottomrule
\end{tabular}
\caption{CER(\%) and WER(\%) on the CV3-Eval Cross-Lingual Zero-Shot subset. The column group indicates the target language, while the sub-column indicates the prompt language.}
\label{tab:cv3_eval_cross_lingual_zeroshot_prompt_lang}
\end{table*}

%% file: tables/ming-freeform-audio-edit-en-full.tex
\begin{table*}[t]
\centering
\small
\renewcommand{\arraystretch}{0.98}
\resizebox{\textwidth}{!}{
\begin{tabular}{llcccc}
\toprule
Edit Type & Model & \multicolumn{4}{c}{Performance} \\
\midrule
\perfhead
\multirow{9}{*}{Insertion}
& Ground Truth   & \pair{--}{--}       & \pair{--}{--}       & \pair{2.99}{3.01} & \pair{--}{--} \\
& VoiceCraft     & \pair{4.39}{4.51} & \pair{0.85}{0.85} & \pair{3.01}{3.03} & \pair{--}{--} \\
& VoiceCraft-X   & \pair{5.74}{6.27} & \pair{0.85}{0.84} & \pair{3.05}{3.06} & \pair{0.166}{0.162} \\
& LEMAS-Edit     & \pair{6.04}{5.10} & \pair{0.81}{0.81} & \pair{3.05}{3.04} & \pair{0.158}{0.171} \\
& ECPA$^\dagger$           & \pair{4.50}{4.97} & \pair{0.82}{0.82} & \pair{3.17}{3.18} & \pair{--}{--} \\
& SSR-Speech     & \pair{\textbf{1.75}}{2.03} & \pair{\textbf{0.94}}{\textbf{0.94}} & \pair{3.06}{3.06} & \pair{0.139}{0.128} \\
& Ming-UniAudio  & \pair{6.49}{7.84} & \pair{0.80}{0.80} & \pair{3.04}{3.04} & \pair{0.168}{0.167} \\
& CosyEdit       & \pair{2.83}{2.85} & \pair{0.86}{0.86} & \pair{3.10}{3.11} & \pair{0.167}{0.167} \\
& CosyEdit2      & \pair{1.90}{\textbf{1.93}} & \pair{0.93}{0.93} & \pair{3.02}{3.03} & \pair{\textbf{0.107}}{\textbf{0.108}} \\
\midrule

\perfhead
\multirow{9}{*}{Deletion}
& Ground Truth   & \pair{--}{--}         & \pair{--}{--}       & \pair{3.05}{3.03} & \pair{--}{--} \\
& VoiceCraft     & \pair{6.07}{5.86}   & \pair{0.81}{0.81} & \pair{3.01}{3.00} & \pair{--}{--} \\
& VoiceCraft-X   & \pair{9.71}{10.65}  & \pair{0.80}{0.80} & \pair{3.01}{3.00} & \pair{0.172}{0.173} \\
& LEMAS-Edit     & \pair{9.29}{8.91}   & \pair{0.80}{0.81} & \pair{3.03}{3.02} & \pair{0.167}{0.169} \\
& ECPA$^\dagger$           & \pair{6.91}{6.88}   & \pair{0.77}{0.78} & \pair{3.09}{3.09} & \pair{--}{--} \\
& SSR-Speech     & \pair{\textbf{5.22}}{\textbf{5.29}}   & \pair{\textbf{0.91}}{\textbf{0.91}} & \pair{3.03}{3.02} & \pair{0.132}{0.134} \\
& Ming-UniAudio  & \pair{14.79}{24.37} & \pair{0.77}{0.75} & \pair{2.97}{2.97} & \pair{0.206}{0.204} \\
& CosyEdit       & \pair{5.69}{5.95}   & \pair{0.83}{0.83} & \pair{3.10}{3.09} & \pair{0.161}{0.164} \\
& CosyEdit2      & \pair{5.52}{5.83}   & \pair{0.90}{0.90} & \pair{3.01}{3.00} & \pair{\textbf{0.131}}{\textbf{0.131}} \\
\midrule

\perfhead
\multirow{9}{*}{Substitution}
& Ground Truth   & \pair{--}{--}       & \pair{--}{--}       & \pair{3.04}{3.05} & \pair{--}{--} \\
& VoiceCraft     & \pair{3.13}{2.96} & \pair{0.80}{0.81} & \pair{3.02}{3.02} & \pair{--}{--} \\
& VoiceCraft-X   & \pair{7.29}{6.73} & \pair{0.81}{0.81} & \pair{3.04}{3.05} & \pair{0.155}{0.146} \\
& LEMAS-Edit     & \pair{6.09}{6.56} & \pair{0.80}{0.80} & \pair{3.05}{3.05} & \pair{0.167}{0.161} \\
& ECPA$^\dagger$           & \pair{4.13}{4.41} & \pair{0.78}{0.78} & \pair{3.15}{3.11} & \pair{--}{--} \\
& SSR-Speech     & \pair{1.90}{1.95} & \pair{\textbf{0.89}}{\textbf{0.90}} & \pair{3.08}{3.08} & \pair{0.146}{0.140} \\
& Ming-UniAudio  & \pair{8.10}{7.95} & \pair{0.77}{0.77} & \pair{3.00}{3.03} & \pair{0.166}{0.178} \\
& CosyEdit       & \pair{2.61}{2.56} & \pair{0.83}{0.83} & \pair{3.11}{3.13} & \pair{0.151}{0.146} \\
& CosyEdit2      & \pair{\textbf{1.43}}{\textbf{1.52}} & \pair{\textbf{0.89}}{\textbf{0.90}} & \pair{3.04}{3.05} & \pair{\textbf{0.137}}{\textbf{0.132}} \\
\bottomrule
\end{tabular}
}
\caption{Full results of performance comparison on the English subset of 
Ming-Freeform-Audio-Edit (extended version of Table~\ref{tab:editing_results_en_dnsmos_mae_split} in the main paper). 
MAE denotes $\mathrm{MAE}_{\mathrm{DNSMOS}}$ between generated and original 
speech, where lower is better. $^\dagger$ECPA refers to 
\citet{ren2026edit} (``Edit Content, Preserve Acoustics''), an abbreviation 
we adopt for brevity as the authors provide no official acronym; results are 
taken directly from the original paper as the model is not publicly released.}
\label{tab:editing_results_en_dnsmos_mae_split_full}
\end{table*}

%% file: tables/ming-freeform-audio-edit-zh.tex
\begin{table*}[t]
\centering
\small
\setlength{\tabcolsep}{4.5pt}
\renewcommand{\arraystretch}{0.98}
\resizebox{\textwidth}{!}{
\begin{tabular}{llcccc}
\toprule
Edit Type & Model & \multicolumn{4}{c}{Performance} \\
\midrule
\perfhead
\multirow{5}{*}{Insertion}
& Ground Truth   & \pair{--}{--}       & \pair{--}{--}       & \pair{3.19}{3.18} & \pair{--}{--} \\
& VoiceCraft-X   & \pair{3.77}{3.57} & \pair{0.84}{0.84} & \pair{3.15}{3.13} & \pair{0.128}{0.138} \\
& LEMAS-Edit     & \pair{6.21}{6.75} & \pair{0.81}{0.82} & \pair{3.17}{3.17} & \pair{0.130}{0.139} \\
& Ming-UniAudio  & \pair{3.73}{3.71} & \pair{0.83}{0.83} & \pair{3.16}{3.17} & \pair{0.137}{0.126} \\
& CosyEdit2      & \pair{\textbf{1.38}}{\textbf{1.33}} & \pair{\textbf{0.89}}{\textbf{0.89}} & \pair{3.20}{3.19} & \pair{\textbf{0.113}}{\textbf{0.114}} \\
\midrule

\perfhead
\multirow{5}{*}{Deletion}
& Ground Truth   & \pair{--}{--}         & \pair{--}{--}       & \pair{3.22}{3.16} & \pair{--}{--} \\
& VoiceCraft-X   & \pair{13.62}{7.18}  & \pair{0.77}{0.79} & \pair{3.04}{3.04} & \pair{0.203}{0.170} \\
& LEMAS-Edit     & \pair{17.43}{8.57}  & \pair{0.75}{0.77} & \pair{3.15}{3.09} & \pair{\textbf{0.159}}{\textbf{0.173}} \\
& Ming-UniAudio  & \pair{12.44}{23.04} & \pair{0.79}{0.81} & \pair{3.07}{3.05} & \pair{0.194}{0.187} \\
& CosyEdit2      & \pair{\textbf{8.23}}{\textbf{3.96}}   & \pair{\textbf{0.85}}{\textbf{0.86}} & \pair{3.09}{3.04} & \pair{0.166}{0.177} \\
\midrule

\perfhead
\multirow{5}{*}{Substitution}
& Ground Truth   & \pair{--}{--}       & \pair{--}{--}       & \pair{3.16}{3.16} & \pair{--}{--} \\
& VoiceCraft-X   & \pair{4.21}{3.88} & \pair{0.82}{0.83} & \pair{3.07}{3.09} & \pair{0.158}{0.152} \\
& LEMAS-Edit     & \pair{8.16}{7.86} & \pair{0.82}{0.81} & \pair{3.10}{3.11} & \pair{0.157}{0.164} \\
& Ming-UniAudio  & \pair{4.76}{4.89} & \pair{0.83}{0.83} & \pair{3.08}{3.08} & \pair{0.161}{0.162} \\
& CosyEdit2      & \pair{\textbf{1.26}}{\textbf{1.39}} & \pair{\textbf{0.89}}{\textbf{0.89}} & \pair{3.13}{3.13} & \pair{\textbf{0.110}}{\textbf{0.125}} \\
\bottomrule
\end{tabular}
}
\caption{Performance comparison on the Chinese subset of Ming-Freeform-Audio-Edit. 
MAE denotes $\mathrm{MAE}_{\mathrm{DNSMOS}}$ between generated and original speech, where lower is better.}
\label{tab:editing_results_zh_dnsmos_mae_split}
\end{table*}

%% file: tables/realedit.tex
\begin{table*}[t]
\centering
\small
\renewcommand{\arraystretch}{0.98}
\begin{tabular}{lccccc}
\toprule
Method & WER $\downarrow$ & SS $\uparrow$ & MCD $\downarrow$ 
& MOS & MAE $\downarrow$ \\
\midrule
GroundTruth    & 6.06 & --    & --   & 3.34 & --   \\
\midrule
FluentSpeech   & 5.97 & 92.74 & --   & 2.72 & 0.78 \\
VoiceCraft     & 6.55 & 97.12 & --   & 3.18 & 0.24 \\
SSR-Speech     & 5.05 & \textbf{98.31} & --   & 3.32 & \textbf{0.14} \\
\noalign{\vskip 2pt}
\hdashline
\noalign{\vskip 2pt}
Ming-UniAudio  & 9.98 & 96.70 & 5.36 & 3.13 & 0.33 \\
CosyEdit       & 4.50 & 97.34 & 4.94 & 3.19 & 0.29 \\
\midrule
CosyEdit2      & \textbf{4.31} & \textbf{97.91} & \textbf{3.93} & 3.21 & \textbf{0.25} \\
\bottomrule
\end{tabular}
\caption{Results for speech editing on RealEdit. 
The dashed line separates cascaded speech editing systems (above) from 
end-to-end models (below). MOS denotes MOSNet-predicted scores; 
MAE denotes $\mathrm{MAE}_{\mathrm{MOSNet}}$ between generated and 
original speech.}
\label{tab:realedit_results}
\end{table*}

%% file: tables/seed-tts-eval.tex
\begin{table}[t]
\centering
\small
\renewcommand{\arraystretch}{0.98}
\begin{tabular}{lcccc}
\toprule
\multirow{2}{*}{Model} 
& \multicolumn{2}{c}{test-zh} 
& \multicolumn{2}{c}{test-en} \\
\cmidrule(lr){2-3} \cmidrule(lr){4-5}
& CER $\downarrow$ & SS $\uparrow$ 
& WER $\downarrow$ & SS $\uparrow$ \\
\midrule
VoiceCraft-X       & 3.26  & 67.6 & 5.13 & 52.7 \\
SSR-Speech         & 20.51 & 53.9 & 9.03 & 48.0 \\
CosyEdit           & 1.76  & 74.7 & 2.98 & 63.5 \\
CosyVoice2         & 1.36  & 75.1 & 3.10 & \textbf{65.7} \\
\midrule
CosyEdit2          & \textbf{1.16} & \textbf{75.2} & \textbf{1.95} & 64.3 \\
\bottomrule
\end{tabular}
\caption{CER(\%), WER(\%) and Speaker Similarity (SS, \%) on Seed-TTS-eval Benchmark.}
\label{tab:seed_eval_results}
\end{table}

%% file: tables/vocoder_reconstruction.tex
\begin{table*}[t]
\centering
\small
\renewcommand{\arraystretch}{0.98}
\setlength{\tabcolsep}{8pt}
\begin{tabular}{llcccccc}
\toprule
Source & Vocoder
  & MR-STFT $\downarrow$ 
  & PESQ $\uparrow$ 
  & STOI $\uparrow$ 
  & ESTOI $\uparrow$ 
  & MCD $\downarrow$ 
  & $\mathrm{MAE}_{\mathrm{DNSMOS}}$ $\downarrow$ \\
\midrule
\multirow{2}{*}{Clean}
  & HiFT-GAN & 1.215 & 3.475 & 0.974 & 0.920 & 1.631 & 0.063 \\
  & BigVGAN  & \textbf{1.138} & \textbf{3.668} & \textbf{0.980} & \textbf{0.936} & \textbf{1.310} & \textbf{0.056} \\
\midrule
\multirow{2}{*}{Noisy}
  & HiFT-GAN & 1.490 & 3.019 & 0.933 & 0.859 & 1.988 & 0.115 \\
  & BigVGAN  & \textbf{1.438} & \textbf{3.185} & \textbf{0.945} & \textbf{0.881} & \textbf{1.630} & \textbf{0.114} \\
\bottomrule
\end{tabular}
\caption{
  Vocoder reconstruction quality on the VoiceBank-DEMAND test set.
  \textit{Source} denotes the subset of the VoiceBank-DEMAND dataset used as vocoder input, either clean or noisy speech.
  Reference-based metrics (MRSTFT, PESQ, STOI, ESTOI, MCD) compare generated audio
  against the source waveform.
  $\mathrm{MAE}_{\mathrm{DNSMOS}}$ is the mean absolute error between DNSMOS scores of generated and source audio.
}
\label{tab:vocoder_reconstruction}
\end{table*}

%% file: tables/reconstruction-fidelity-experiment.tex
\begin{table}[t]
\centering
\small
\renewcommand{\arraystretch}{0.98}
\setlength{\tabcolsep}{16pt}
\begin{tabular}{lcc}
\toprule
Method & SS $\uparrow$ & MCD $\downarrow$ \\
\midrule
HiFT-GAN$^\dagger$  & 99.02 & 3.03 \\
BigVGAN$^\dagger$   & \textbf{99.25} & \textbf{2.81} \\
\midrule
CosyVoice2  & 96.92 & 6.24 \\
CosyEdit2   & \textbf{99.08} & \textbf{3.07} \\
\bottomrule
\end{tabular}
\caption{
  Speaker similarity (SS, \%) and mel-cepstral distortion (MCD, dB) on the speech preservation evaluation over RealEdit, where the target text is identical to the prompt/original text. $^\dagger$Oracle vocoder reconstruction upper bounds, where the mel-spectrogram is extracted directly from the prompt/original speech.
}
\label{tab:reconstruction-fidelity-experiment}
\end{table}

%% file: tables/ming-en-subjective-evaluation.tex
\begin{table*}[t]
\centering
\small
\renewcommand{\arraystretch}{0.98}
\setlength{\tabcolsep}{12pt}
\begin{tabular}{llccc}
\toprule
Edit Type & Model & \multicolumn{3}{c}{Subjective Evaluation} \\
\midrule
\subjhead
\multirow{5}{*}{Insertion}
& VoiceCraft    & 4.353 & 4.497 & 4.143 \\
& SSR-Speech    & 4.550 & \textbf{4.637} & \textbf{4.247} \\
& Ming-UniAudio & 4.213 & 4.373 & 3.937 \\
& CosyEdit      & 4.543 & 4.587 & 4.150 \\
& CosyVoice2    & 4.650 & 4.473 & 3.690 \\
& CosyEdit2     & \textbf{4.773} & \textbf{4.637} & 4.213 \\
\midrule

\subjhead
\multirow{5}{*}{Deletion}
& VoiceCraft    & 4.377 & 4.627 & 4.373 \\
& SSR-Speech    & 4.597 & 4.647 & \textbf{4.407} \\
& Ming-UniAudio & 3.057 & 4.413 & 3.140 \\
& CosyEdit      & 4.553 & 4.657 & 4.220 \\
& CosyVoice2    & 4.623 & 4.563 & 3.927 \\
& CosyEdit2     & \textbf{4.660} & \textbf{4.777} & 4.403 \\
\midrule

\subjhead
\multirow{5}{*}{Substitution}
& VoiceCraft    & 4.443 & 4.617 & 4.370 \\
& SSR-Speech    & 4.760 & 4.683 & 4.513 \\
& Ming-UniAudio & 4.283 & 4.507 & 3.990 \\
& CosyEdit      & 4.563 & 4.703 & 4.370 \\
& CosyVoice2    & 4.777 & 4.653 & 4.087 \\
& CosyEdit2     & \textbf{4.797} & \textbf{4.730} & \textbf{4.520} \\
\bottomrule
\end{tabular}
\caption{Subjective evaluation results on the English subset of Ming-Freeform-Audio-Edit. 
IMOS, SMOS, and PMOS denote intelligibility, speaker similarity, and preservation mean opinion scores, respectively.}
\label{tab:subjective_results_en}
\end{table*}

%% file: tables/ming-zh-subjective-evaluation.tex
\begin{table*}[t]
\centering
\small
\renewcommand{\arraystretch}{0.98}
\setlength{\tabcolsep}{12pt}
\begin{tabular}{llccc}
\toprule
Edit Type & Model & \multicolumn{3}{c}{Subjective Evaluation} \\
\midrule
\subjhead
\multirow{4}{*}{Insertion}
& VoiceCraft-X  & 4.373 & 4.570 & 3.707 \\
& LEMAS-Edit    & 4.247 & 4.370 & 3.673 \\
& Ming-UniAudio & 4.333 & 4.623 & 4.107 \\
& CosyVoice2    & 4.900 & 4.553 & 4.037 \\
& CosyEdit2     & \textbf{4.907} & \textbf{4.767} & \textbf{4.500} \\
\midrule

\subjhead
\multirow{4}{*}{Deletion}
& VoiceCraft-X  & 4.313 & 4.727 & 3.963 \\
& LEMAS-Edit    & 4.080 & 4.663 & 3.897 \\
& Ming-UniAudio & 3.307 & 4.567 & 3.173 \\
& CosyVoice2    & 4.713 & 4.573 & 3.867 \\
& CosyEdit2     & \textbf{4.720} & \textbf{4.863} & \textbf{4.303} \\
\midrule

\subjhead
\multirow{4}{*}{Substitution}
& VoiceCraft-X  & 4.297 & 4.647 & 3.520 \\
& LEMAS-Edit    & 4.007 & 4.453 & 3.630 \\
& Ming-UniAudio & 4.250 & 4.677 & 3.720 \\
& CosyVoice2    & 4.863 & 4.617 & 4.090 \\
& CosyEdit2     & \textbf{4.883} & \textbf{4.857} & \textbf{4.527} \\
\bottomrule
\end{tabular}
\caption{Subjective evaluation results on the Chinese subset of Ming-Freeform-Audio-Edit. 
IMOS, SMOS, and PMOS denote intelligibility, speaker similarity, and preservation mean opinion scores, respectively.}
\label{tab:subjective_results_zh}
\end{table*}

%% file: tables/cv3-eval-subjective-evaluation.tex
\begin{table}[t]
\centering
\small
\renewcommand{\arraystretch}{0.98}
\setlength{\tabcolsep}{7pt}
\begin{tabular}{llcc}
\toprule
Language & Model & IMOS$\uparrow$ & SMOS$\uparrow$ \\
\midrule
\multirow{4}{*}{English}
& VoiceCraft-X & 3.675 & 3.418 \\
& CosyEdit     & 4.628 & 4.305 \\
& CosyVoice2   & 4.658 & 4.342 \\
& CosyEdit2    & \textbf{4.730} & \textbf{4.418} \\
\midrule
\multirow{4}{*}{Chinese}
& VoiceCraft-X & 3.882 & 3.530 \\
& CosyEdit     & 4.275 & 4.272 \\
& CosyVoice2   & 4.535 & 4.432 \\
& CosyEdit2    & \textbf{4.640} & \textbf{4.512} \\
\bottomrule
\end{tabular}
\caption{Subjective evaluation results for zero-shot TTS on the CV3-Eval Multi-lingual Voice Cloning subset. 
IMOS and SMOS denote intelligibility and speaker similarity mean opinion scores, respectively.}
\label{tab:zeroshot_tts_subjective}
\end{table}